# Left-Right Symmetry at LHC

Alessio Maiezza,[1] Miha Nemevšek,[2] Fabrizio Nesti,[3] Goran Senjanović[4]

[1] *Dipartimento di Fisica, Università dell'Aquila, v. Vetoio, I-67010, L'Aquila, and INFN – LNGS, Assergi, I-67010, L'Aquila, Italy*

[2] *II. Institut für Theoretische Physik, Universität Hamburg, Luruper Chaussee 149, 22761 Hamburg, Germany, and*
*Jožef Stefan Institute, 1000 Ljubljana, Slovenia*

[3] *Dipartimento di Fisica, Università di Ferrara, v. Saragat 1, I-44100, Ferrara, Italy*

[4] *ICTP, strada Costiera 11, I-34151, Trieste, Italy*

**Abstract**

We revisit the issue of the limit on the scale of Left-Right symmetry breaking. We focus on the minimal $SU(2)_L \times SU(2)_R \times U(1)_{B-L}$ gauge theory with the seesaw mechanism and discuss the two possibilities of defining Left-Right symmetry as parity or charge conjugation. In the commonly adopted case of parity, we perform a complete numerical study of the quark mass matrices and the associated left and right mixing matrices without any assumptions usually made in the literature about the ratio of vacuum expectation values. We find that the usual lower limit on the mass of the right-handed gauge boson from the K mass difference, $M_{W_R} > 2.5\,\text{TeV}$, is subject to a possible small reduction due to the difference between right and left Cabibbo angles. In the case of charge conjugation the limit on $M_{W_R}$ is somewhat more robust. However, the more severe bounds from CP-violating observables are absent in this case. In fact, the free phases can also resolve the present mild discrepancy between the Standard Model and CP-violation in the *B*-sector. Thus, even in the minimal case, both charged and neutral gauge bosons may be accessible at the Large Hadron Collider with spectacular signatures of lepton number violation.

# Contents

## 1 Introduction

The most striking and probably the least appealing signature of the standard model (SM) is a complete asymmetry between left and right. This is simply a reflection of the chiral structure of weak interactions, a feature that sets it apart from the rest of the forces in nature and a feature that has been puzzling us for more than a half a century. It is interesting to note that in the original work on the breakdown of parity [1], Lee and Yang discussed at length the possibility of restoring parity at high energies, through the existence of mirror fermions.

An alternative is provided by Left-Right (LR) symmetric theories, where the standard weak interactions remain chiral, but a $W$ boson of the SM obtains its mirror-like twin. In simple terms, instead of mirror fermions one has here mirror gauge bosons. This is certainly more economical, for instead of doubling the whole fermion spectrum, one only doubles the weak (charged and neutral) gauge bosons. The minimal L-R symmetric model [2, 3] is thus based on the gauge group $SU(2)_L \times SU(2)_R \times U(1)_{B-L}$, completed by a symmetry between the two gauge $SU(2)$ groups. This LR symmetry is then shown to be broken spontaneously [3].

The choice of LR symmetry is twofold: (i) a generalized parity $\mathcal{P}$ and (ii) a generalized charge conjugation $\mathcal{C}$. The case (i) of $\mathcal{P}$ was used originally and is commonly adopted, but the case (ii) of $\mathcal{C}$ should be considered equally, if not more, motivated for being an automatic gauge symmetry in the $SO(10)$ grand unified theory. Either of these discrete symmetries plays an important role in relating the couplings of the theory, especially the Yukawa ones. In the former case the Yukawa matrices are hermitian while in the latter case they become symmetric. This brings important restrictions on the pattern of the left and right quark and lepton mixing matrices.

We carefully study both cases. Our principal interest and the central aspect of this paper is to determine the precise lower limit on the Left-Right symmetric scale in these minimal models, conventionally identified with the mass of the right-handed charged gauge boson $W_R$. In fact, while a high scale of restoration of Left-Right symmetry fits well within the framework of grand unification and neutrino masses [4] this scale may be as low as TeV and thus can be very interesting in view of the forthcoming experimental probe of the TeV energy range at the Large Hadron Collider (LHC). Since clearly $W_R$ cannot be arbitrarily light, in view of the chiral nature of the SM, it is crucial to know what the precise limit on it mass is and whether it is truly accessible at the LHC.

The strongest limit on $M_{W_R}$ comes from the neutral $K$-meson system, which together with the $B$-system was the subject of a dedicated effort, starting from the original work [5], over the last thirty years [6, 7, 8, 9, 10, 11, 12, 13]. The lower bound depends crucially on the analog of the Cabibbo-Kobayashi-Maskawa (CKM) mixing matrix in the Right sector, $V_R$. It was in fact realized [14] that in non-minimal models, where $V_R$ may be very different from the CKM matrix, the bounds can be relaxed and $M_{W_R}$ is only limited by direct searches.

In minimal models on the other hand where the left and right CKM matrices are related, there is a lower bound in the TeV range. For instance, for equal mixing angles, the original bound $M_{W_R} > 1.6\,\text{TeV}$ from the $K_L$-$K_S$ mass difference [5] has grown over the years to $M_{W_R} > 2.5\,\text{TeV}$ [15, 12]. In the minimal model the masses of the neutral and charged heavy gauge bosons are connected as $M_{Z_R} \simeq 1.7 M_{W_R}$ and thus $M_{Z_R} > 4\,\text{TeV}$. Although quite large, these values are still compatible with the LHC discovery of the LR symmetry, as we discuss in section 4.1.

It was further argued that the bounds from CP-violating observables are much stronger [12] and would push the $W_R$ scale beyond the LHC reach. These bounds disappear completely in the case (ii) with Left-Right symmetry as charge conjugation $\mathcal{C}$. This is the central new result of our paper and will be discussed at length. It can provide a strong impetus to consider $\mathcal{C}$ as an underlying LR symmetry.

In the case of parity being LR symmetry, it is still important to assess the magnitude of the right mixing angles and phases, in order to know how robust the limits are. Lowering them could be essential in the early stage of LHC, before one reaches the planned energy of 14 TeV. This requires a complete study of the quark mass matrices without making any assumptions on the relevant vacuum expectation values (VEV) as normally done in the literature. This is what we have done in the conventional case of $\mathcal{P}$ as a LR symmetry. Our study shows that indeed the left and right mixing matrices are not necessarily the same, and the bounds cited in the literature [12] can be lowered by 10–20% (at best) exploiting the uncertainties in quark masses and mixing angles.

As we will discuss throughout our paper, the only true phenomenological limit on the LR scale in the minimal theory is $M_{W_R} \gtrsim 2\text{–}3\,\text{TeV}$, as long as LR symmetry is chosen to be charge conjugation. The scale of LR symmetry restoration is thus perfectly accessible at LHC, even in the early stage. As envisioned in [16], it would have spectacular signatures of lepton-number violation in the form of same-sign dileptons and it could in principle allow the determination of both the $W_R$ and right-handed neutrino mass.

The paper is organized as follows, in section 2 we define the minimal models. The main section of our paper is 3, where we offer an in-depth analysis of left and right CKM matrices and where we derive the bounds on the LR scale. In section 4 we discuss the LHC reach and briefly comment on the neutrinoless double beta decay ($0\nu\beta\beta$) as well as lepton flavour violation (LFV). Finally in section 5 we give our conclusions and outlook. The technical details are left for the Appendix.

## 2 LR Models: $\mathcal{P}$ vs $\mathcal{C}$

The minimal L-R symmetric theory is based on the following gauge group (suppressing colour):

$$G_{LR} = SU(2)_L \times SU(2)_R \times U(1)_{B-L},$$

plus a symmetry between the left and right sectors. Quarks and leptons are completely LR symmetric

$$Q_{L,R} = \begin{pmatrix} u \\ d \end{pmatrix}_{L,R}, \qquad \ell_{L,R} = \begin{pmatrix} \nu \\ e \end{pmatrix}_{L,R}. \tag{1}$$

The formula for the electromagnetic charge becomes

$$Q_{em} = I_{3L} + I_{3R} + \frac{B-L}{2}. \tag{2}$$

The Higgs sector consists of the following multiplets [17]: the bi-doublet $\Phi \in (2_L, 2_R, 0)$ and the $SU(2)_{L,R}$ triplets $\Delta_L \in (3_L, 1_R, 2)$ and $\Delta_R \in (1_L, 3_R, 2)$, according to the $SU(2)_L \times SU(2)_R \times U(1)_{B-L}$ quantum numbers

$$\Phi = \begin{bmatrix} \phi_1^0 & \phi_2^+ \\ \phi_1^- & \phi_2^0 \end{bmatrix} \qquad \Delta_{L,R} = \begin{bmatrix} \Delta^+/\sqrt{2} & \Delta^{++} \\ \Delta^0 & -\Delta^+/\sqrt{2} \end{bmatrix}_{L,R} \tag{3}$$

It can be shown that the first stage of the breaking of the $G_{LR}$ down to the SM model symmetry, takes the following form [17]

$$\langle \Delta_L \rangle = 0 \quad , \quad \langle \Delta_R \rangle = \begin{bmatrix} 0 & 0 \\ v_R & 0 \end{bmatrix} \tag{4}$$

At the next stage, the neutral components $\Phi$ develop a VEV and break the SM symmetry down to $U(1)_{em}$

$$\langle \Phi \rangle = \begin{bmatrix} v_1 & 0 \\ 0 & v_2 \, e^{i\alpha} \end{bmatrix} \tag{5}$$

where $v_{1,2}$ are real and positive, $M_W^2 = g^2 v^2 \equiv g^2(v_1^2 + v_2^2)$ and $g \equiv g_L = g_R$ denote the SU(2) gauge couplings. In turn, $\Delta_L$ develops a tiny VEV $\langle \Delta_L \rangle \propto v^2/v_R$.

**Gauge bosons.** The gauge boson masses are given by

$$M_{W_R}^2 \simeq g^2 \, v_R^2 \tag{6}$$

$$M_{Z_R}^2 \simeq 2(g^2 + g_{B-L}^2)\, v_R^2 = \frac{2g^2}{g^2 - g_Y^2} M_{W_R}^2 \simeq 3 M_{W_R}^2, \tag{7}$$

where we used the relation $g_Y^{-2} = g^{-2} + g_{B-L}^{-2}$ among $g_Y$ and $g_{B-L}$, respectively the gauge couplings of $Y/2$ and $(B-L)/2$. In other words, $M_{Z_R} \simeq 1.7 M_{W_R}$ and the limit on $W_R$ becomes even more important if one wishes to discover also $Z_R$ at LHC.

In the above, we neglected the mixing among left and right gauge bosons which comes from the product of the VEVs $v_1$ and $v_2$. Although in general these mixings could play an important role, for very large LR scale they obviously become secondary or irrelevant. In other words this will be justified by the largeness of the LR scale which will emerge from the phenomenological analysis. In fact, it is also worth pointing out that for the same reason the phenomenological limits on the $W_L$-$W_R$ mixing do not require anymore the smallness of the VEV ratio $v_1/v_2$. When necessary we will nevertheless consider also this tiny mixing and comment on its consequences.

**Yukawa sector.** We will be mostly interested in the quark mixing matrices, crucial for the study of K and B meson systems. The quark Yukawa interaction is

$$\mathcal{L}_Y = \overline{Q}_L (Y\, \Phi + \tilde{Y}\, \tilde{\Phi}) Q_R + h.c., \tag{8}$$

where $\tilde{\Phi} \equiv \sigma_2 \Phi^* \sigma_2$ and we suppress the generation indices. The quark mass matrices are then

$$\begin{aligned} M_u &= v\, (Y\, c + \tilde{Y}\, s\, e^{-i\alpha}) \\ M_d &= v\, (Y\, s\, e^{i\alpha} + \tilde{Y}\, c), \end{aligned} \tag{9}$$

where $s = v_2/v$, $c = v_1/v$, and below it will be convenient to use $x = v_2/v_1$, with $0 < x < 1$.

As usual one diagonalizes the mass matrices

$$M_u = U_{uL}\, m_u\, U_{uR}^\dagger, \qquad M_d = U_{dL}\, m_d\, U_{dR}^\dagger, \tag{10}$$

with the net result of flavour changing charged weak interactions

$$\mathcal{L}_{CC} = \frac{g}{\sqrt{2}} \left[ W_L^\mu \begin{pmatrix} \bar{u} & \bar{c} & \bar{t} \end{pmatrix}_L V_L \gamma_\mu \begin{pmatrix} d \\ s \\ b \end{pmatrix}_L + W_R^\mu \begin{pmatrix} \bar{u} & \bar{c} & \bar{t} \end{pmatrix}_R V_R \gamma_\mu \begin{pmatrix} d \\ s \\ b \end{pmatrix}_R \right] + \text{h.c.}, \quad (11)$$

where $V_L = U_{uL}^\dagger U_{dL} = V_L^{CKM}$ is the CKM matrix in the canonical form, and $V_R = U_{uL}^\dagger U_{dL}$ its right-handed analogue, which has in principle different angles and five extra phases. We can extract the extra phases from $V_R$ and write it as

$$V_R = K_u V_R^{CKM} K_d, \quad (12)$$

where now $V_R^{CKM}$ is the analogue of $V_L^{CKM}$ and $K_{u,d}$ are diagonal matrices of phases

$$K_u = \text{diag}(e^{i\theta_u}, e^{i\theta_c}, e^{i\theta_t}), \qquad K_d = \text{diag}(e^{i\theta_d}, e^{i\theta_s}, e^{i\theta_b}), \quad (13)$$

and one linear combination can be set to zero.

**Heavy Higgs.** In the quark sector only the bidoublet matters. It contains two $SU(2)_L$ doublets and its form is in general messy. While one doublet must be made light (at the weak scale) the other gets a large mass proportional to the LR breaking scale. The reason why it must be heavy is its neutral flavour violating interactions. As it is known, these force its mass to be even higher than $m_{W_R}$ and thus it can be considered as a degenerate multiplet of $SU(2)_L$, since its mass split is of order of the weak scale. This simplifies the form of its interactions, which can be written as

$$\mathcal{L}_H = \bar{Q}_L \left[ \frac{M_d - 2cse^{-i\alpha} M_u}{\sqrt{2}v(c^2 - s^2)} \right] H u_R + \bar{Q}_L \left[ \frac{M_u - 2cse^{i\alpha} M_d}{\sqrt{2}v(c^2 - s^2)} \right] \tilde{H} d_R + h.c. \quad (14)$$

where $H$ is the heavy doublet. Its neutral component mediates FCNCs via the above couplings. Therefore, the dangerous FCNC coupling for 'down' quarks stem from the mass matrix of the 'up' quarks, and vice-versa. It is also worth noting that one can not have enhancements from the denominators, by the requirement of having perturbative Yukawa couplings (the ratio $s/c = x$ should not exceed $\sim 0.8$). The interesting FCNC part can then be written conveniently in the mass basis as

$$\mathcal{L}_H^{FCNC} \simeq \frac{g}{2M_{W_L}} \left[ \bar{u}_L \left( V_L m_d V_R^\dagger \right) u_R H_0 + \bar{d}_L \left( V_L^\dagger m_u V_R \right) d_R H_0^* \right] + h.c. \quad (15)$$

where $H_0$ is the canonically normalized (complex) neutral component of $H$.

It will be useful for later purpose to be more specific regarding the heavy Higgs mass: $m_H^2 = \lambda_H v_R^2$ where $\lambda_H$ is an appropriate quartic coupling in the Higgs potential (see table I in [12]). We will ask from the theory to remain perturbative, analogously to the Higgs perturbativity limit in the SM, $M_h < 10\, M_W \simeq 800\, \text{GeV}$. In the LR model, in the absence of a precise analogous assessment in the LR model we will stick to a similar bound, $M_H < 10\, M_{W_R}$.

**LR symmetries.** The pattern of mass matrices depends on the kind of Left-Right symmetry imposed on the model in the high-energy, symmetric phase. It is easy to verify that the only realistic discrete symmetries exchanging the left and right sectors, preserving the kinetic terms are[1]

$$\mathcal{P}: \begin{cases} Q_L \leftrightarrow Q_R \\ \Phi \to \Phi^\dagger \end{cases} \qquad \mathcal{C}: \begin{cases} Q_L \leftrightarrow (Q_R)^c \\ \Phi \to \Phi^T \end{cases} \qquad (16)$$

where $(Q_R)^c = C\gamma_0 Q_R^*$ is the charge-conjugate spinor.

The names of $\mathcal{P}$ and $\mathcal{C}$ are motivated by the fact that they are directly related to parity and charge conjugation supplemented by the exchange of the left and right SU(2) gauge groups, as is evident from (16).

Note that $(Q_R)^c$ is a spinor of left chirality like $Q_L$, and thus $\mathcal{C}$ has an important advantage: since it involves the spinors with same final chirality, it can be gauged, i.e. it allows to have this symmetry embedded in a local gauge symmetry. In fact, in the SO(10) grand unified theory $\mathcal{C}$ is a finite gauge transformation. The gauging is not only an aesthetic advantage, it guarantees the protection from unknown high energy physics, gravitational effects, etc.

In spite of this, the simpler case of $\mathcal{P}$ was the main subject of past investigations [2], probably for historical reasons since the original papers used it. The case of $\mathcal{C}$ on the other hand was not extensively studied, at least not in the context of phenomenology. For this reason we devote careful attention to this case too.

The LR symmetries $\mathcal{P}$ or $\mathcal{C}$ pose nontrivial restrictions on the mass matrices $M_u$, $M_d$. In fact, the Yukawa term is invariant under either symmetry provided the Yukawa matrices are respectively hermitian or symmetric:

$$\mathcal{P}: Y = Y^\dagger, \qquad \mathcal{C}: Y = Y^T. \qquad (17)$$

**Case of $\mathcal{P}$.** Here, even if the Yukawa matrices are hermitian, the quark masses in general are not, due to the 'spontaneous' phase $\alpha$. This phase, as one can see in equation (9), is effectively suppressed by $v_2/v_1$. Therefore in the limit of vanishing $\alpha$ or $v_2/v_1$ the quark mass matrices become hermitian and

$$\mathcal{P}: V_R \simeq S_u V_L S_d, \qquad (18)$$

where $S_u$, $S_d$ are diagonal matrices of signs or, using the notation of (12), $\theta_{u,c,t,d,s,b} = 0, \pi$.

One may think that for large $v_2/v_1 \sim O(1)$ the impact of $\alpha$ becomes important, and that the L and R mixing matrices might be disentangled. However, a complete analysis shows that this is not the case, because for large $v_2/v_1$ the phase $\alpha$ is forced to be very small to accommodate the known quark masses. We prove this in appendix A in full generality and derive an analytic bound on $\alpha$. As a result, the mass matrices are always approximately hermitian, and the left and right mixing matrices end up being anyway very near. In practice the mixing angles vary at most by 20% (at 95%CL) and, beside the signs, the phase differences are very small, $\lesssim 0.01$.

---
[1] There are in fact only two other possibilities, $\Phi \leftrightarrow \tilde{\Phi}^\dagger$ and $\Phi \leftrightarrow \tilde{\Phi}^T$, that however lead to unrealistically related mass matrices, respectively $M_u = M_d^\dagger$ in the first case and $M_u = M_d^T$ or $M_u = M_d^*$ in the second.

The most important consequence is the near equality of left and right mixing angles which implies no suppression for the $W_R$ production at colliders, if $W_R$ is light enough. For this reason the bound on its mass becomes experimentally important in view of the forthcoming LHC, which prompted a recent detailed study [12]. They find that the bound from CP violating amplitudes would kill any hope of seeing LR symmetry at LHC. Motivated by their claim, we performed a complete analytical and numerical analysis which supports their findings. We point out though that there is no limit form the electric dipole moment (EDM). In the absence of the EDM limit the main bound, resulting from $\epsilon'$, is lowered to roughly $M_{W_R} \gtrsim 3.1\,\text{TeV}$. This would still be reachable at the 14 TeV LHC, as discussed below. On the other hand, $\epsilon'$ is also subject to its own theoretical uncertainties, and we leave to the reader to decide on the relevance of such a bound.

**Case of $\mathcal{C}$.** Via equation (9), the mass matrices are themselves symmetric, with the result that the Left and right mixing matrices are related as

$$\mathcal{C}: \ V_R = K_u V_L^* K_d, \tag{19}$$

where $K_u$, $K_d$ are diagonal matrices of phases as in (13). As a first consequence, the mixing angles are the same, and this is often called 'pseudo-manifest' LR symmetry situation. In addition, there are five new (physical) phases in $K_u$, $K_d$, which are unconstrained. These phases have a direct impact on the CP-violating observables. The remarkable result is that in the limit of these phases vanishing, the LR box diagrams become manifestly real.[2] This fact has a profound consequence on the bound for the LR scale. It means that only CP conserving processes are relevant, and as we discuss at length below, in the $K$-meson system this will give the bound $M_{W_R} \gtrsim 2.5\,\text{TeV}$. The larger bounds from $\epsilon$ and/or $\epsilon'$ disappear completely. The $B_d$, $B_s$ meson systems are completely analogous, the LR contribution to CP-violation can be set to zero. However, the situation is somehow more intriguing, since a slight discrepancy of the SM with data was reported [19, 20] in both the $B_d$ and $B_s$ CP violating phases. This discrepancy is explained very well by the contribution from the LR sector with nonzero phases, for a preferred window of $M_{W_R}$ which is exactly in the energy region addressed by LHC.

## 3 Left-Right symmetry and bounds on New Physics

The relevant bounds come from the neutral meson mixings. We review first the New Physics (NP) contributions to the CP-conserving and CP-violating processes, and then summarize the resulting bounds on the LR scale for the $\mathcal{P}$ and $\mathcal{C}$ cases in the relative sections.

---

[2]This was noticed first in [8], who suggested the symmetric mass matrices in the case of $\mathcal{P}$ with spontaneous CP violation. In this case, the Yukawa couplings are real and symmetric and so the quark mass matrices remain symmetric. Obviously in this case $\mathcal{P}$ and $\mathcal{C}$ become equivalent. It turns out though that spontaneous CP does not work [18], since the second doublet in the bidoublet must be very heavy, as discussed below.

**Hamiltonians.** The analysis concerns mostly the $\Delta F = 2$ processes. The effective hamiltonian contains contributions from the new gauge and Higgs bosons. The gauge boson box diagrams with only $W_L$, mixed $W_L$ and $W_R$, and only $W_R$ gauge boson exchanges are, respectively

$$\mathcal{H}_{LL}^{\Delta F=2} = \frac{G_F^2 M_{W_L}^2}{4\pi^2} \sum_{d,d'=d,s,b} \bar{d}'\gamma_\mu P_L d\, \bar{d}'\gamma_\mu P_L d \sum_{i,j=c,t} \lambda_i^{LL}\lambda_j^{LL} S_{LL}(x_i, x_j)\, \eta_{LL,ij} \qquad (20)$$

$$\mathcal{H}_{LR}^{\Delta F=2} = \frac{G_F^2 M_{W_L}^2}{4\pi^2}\, 8\,\beta \sum_{d,d'=d,s,b} \bar{d}' P_L d\, \bar{d}' P_R d \sum_{i,j=u,c,t} \lambda_i^{LR}\lambda_j^{RL} S_{LR}(x_i, x_j, \beta)\, \eta_{LR,ij} \qquad (21)$$

$$\mathcal{H}_{RR}^{\Delta F=2} = \frac{G_F^2 M_{W_L}^2}{4\pi^2}\, \beta \sum_{d,d'=d,s,b} \bar{d}'\gamma_\mu P_R d\, \bar{d}'\gamma_\mu P_R d \sum_{i,j=c,t} \lambda_i^{RR}\lambda_j^{RR} S_{RR}(x_i, x_j, \beta)\, \eta_{RR,ij}, \qquad (22)$$

where $\lambda_i^{AB} = V_{id'}^{A*} V_{id}^B$, $x_i = (m_i/M_{W_L})^2$, $\beta = M_{W_L}^2/M_{W_R}^2$ and we use the GIM mechanism in the LL and RR diagrams to eliminate $\lambda_u$. The loop functions are then [5, 6, 55]

$$S_{LR}(x_i, x_j, \beta) = \frac{1}{4}\sqrt{x_i x_j}\left[(4 + x_i x_j \beta) I_1(x_i, x_j, \beta) - (1 + \beta) I_2(x_i, x_j, \beta)\right] \qquad (23)$$

$$S_{LL}(x_i, x_j) = S_{RR}(x_i, x_j, 1) \qquad (24)$$

$$S_{RR}(x_i, x_j, \beta) = f_{RR}(x_i, x_j, \beta) - f_{RR}(x_u, x_i, \beta) - f_{RR}(x_u, x_j, \beta) + f_{RR}(x_u, x_u, \beta) \qquad (25)$$

$$f_{RR}(x_i, x_j, \beta) = \frac{1}{4}\left[(4 + x_i x_j) I_2(x_i, x_j, \beta) - 8\, x_i x_j I_1(x_i, x_j, \beta)\right], \qquad (26)$$

with

$$I_1 = \frac{x_i \ln x_i}{(1-x_i)(1-x_i\beta)(x_i - x_j)} + (i \leftrightarrow j) - \frac{\beta \ln \beta}{(1-\beta)(1-x_i\beta)(1-x_j\beta)} \qquad (27)$$

$$I_2 = \frac{x_i^2 \ln x_i}{(1-x_i)(1-x_i\beta)(x_i - x_j)} + (i \leftrightarrow j) - \frac{\ln \beta}{(1-\beta)(1-x_i\beta)(1-x_j\beta)}. \qquad (28)$$

Notice the enhancement factor of 8 in the LR contribution (19), which will play an important role in enhancing the LR contribution in the small $x$ limit.[3]

The FCNC higgs contribution to the $\Delta F = 2$ hamiltonian has the same structure as the one from the LR box,

$$\mathcal{H}_H^{\Delta F=2} = -\frac{4 G_F}{\sqrt{2} M_H^2} \sum_{d,d'=d,s,b} \bar{d}' P_L d\, \bar{d}' P_R d \sum_{i,j=u,c,t} \lambda_i^{LR}\lambda_j^{RL} m_i m_j\,. \qquad (29)$$

Since it is not loop-suppressed it leads to a strong bound on the mass of $H$, as we discuss later.

---

[3]This normalization was chosen historically for the $K$ meson system where $\Delta m_K$ was used to set the large limit on $W_R$. There the charm quark dominates, thus the relevance of small $x$.

Summarizing, the total hamiltonian contains the contributions from the gauge and higgs exchanges:
$$\mathcal{H}_{full}^{\Delta F=2} = \mathcal{H}_{LL}^{\Delta F=2} + \mathcal{H}_{LR}^{\Delta F=2} + \mathcal{H}_{RR}^{\Delta F=2} + \mathcal{H}_{H}^{\Delta F=2}. \qquad (30)$$

For each physical process, it has to be renormalized at the relevant meson scale, which is accomplished by the factors $\eta$ in equations (18)–(20). Since in most cases each loop is dominated by a specific quark exchange (see below) we have $\eta_{LL}^K \equiv \eta_{LL,cc}^K = 1.4$ [9, 21] (while $\eta_{LL,ct}^K = 0.47$, $\eta_{LL,tt}^K = 0.57$ [21]) $\eta_{LL}^{B_{d,s}} \equiv \eta_{LL,tt}^{B_{d,s}} = 0.55$ [22]. In the LR sector $\eta_{LR}^K \equiv \eta_{LR,cc}^K = 1.4$ [9], $\eta_{LR}^{B_{d,s}} \equiv \eta_{LR,tt}^B = 2.12$ [22]. For the $B$ mesons, we will need also the RR contributions, whose renormalization we recalculate as $\eta_{RR}^B = 0.50$.[4] The Higgs FCNC Hamiltonian is not renormalized, since its anomalous dimension is compensated by the running of the quark masses [9].

**Matrix elements.** The matrix elements of the above operators for the $K$, $B_d$, $B_s$ mesons can be computed in chiral perturbation theory and corrected by lattice calculation. The results for the LL, RR and LR contributions are

$$\left\langle M^0 \left| \bar{d}' \gamma_\mu P_L d \, \bar{d}' \gamma_\mu P_L d \right| \overline{M}^0 \right\rangle = \left\langle M^0 \left| \bar{d}' \gamma_\mu P_R d \, \bar{d}' \gamma_\mu P_R d \right| \overline{M}^0 \right\rangle = \frac{2}{3} f_M^2 m_M \mathcal{B}_M^{LL} \qquad (31)$$

$$\left\langle M^0 \left| \bar{d} P_L d' \, \bar{d} P_R d' \right| \overline{M}^0 \right\rangle = \frac{1}{2} f_M^2 m_M \mathcal{B}_M^{LR} \left[ \left( \frac{m_M}{m_{d'} + m_d} \right)^2 + \frac{1}{6} \right], \qquad (32)$$

where $M$ represents either $K$ or $B$ mesons, composed of the $d$ and $d'$ quarks. For the decay constants $f_M$ and the nonperturbative factors $\mathcal{B}_M^{LL}$, we use the averages adopted by the CKM-fitter group [20], obtained from recent determinations: $f_K = 0.113\,\text{GeV}$, $f_{B_d} = 0.134\,\text{GeV}$, $f_{B_s} = 0.161\,\text{GeV}$, and $\mathcal{B}_K^{LL} = 0.721(05)(40)$, $\mathcal{B}_{B_d}^{LL} = 1.17(5)(7)$, $\mathcal{B}_{B_s}^{LL} = 1.23(3)(5)$. For the LR operators we use the recent determinations $\mathcal{B}_K^{LR} = 0.81$ [23], $\mathcal{B}_{B_d}^{LR} = 1.15$, $\mathcal{B}_{B_s}^{LR} = 1.16$ [24, 25].

In order to ease the reader's pain and for the sake of transparency of numerical results we collect the values of the relevant matrix elements:

$$\left\langle K^0 \left| \bar{d} \gamma_\mu P_L s \, \bar{d} \gamma_\mu P_L s \right| \overline{K}^0 \right\rangle \simeq 0.00304\,\text{GeV}^3 \qquad \left\langle K^0 \left| \bar{d} P_L s \, \bar{d} P_R s \right| \overline{K}^0 \right\rangle \simeq 0.059\,\text{GeV}^3 \qquad (33)$$

$$\left\langle B_d \left| \bar{d} \gamma_\mu P_L b \, \bar{d} \gamma_\mu P_L b \right| \overline{B}_d \right\rangle \simeq 0.074\,\text{GeV}^3 \qquad \left\langle B_d \left| \bar{d} P_L b \, \bar{d} P_R b \right| \overline{B}_d \right\rangle \simeq 0.096\,\text{GeV}^3 \qquad (34)$$

$$\left\langle B_s \left| \bar{s} \gamma_\mu P_L b \, \bar{s} \gamma_\mu P_L b \right| \overline{B}_s \right\rangle \simeq 0.114\,\text{GeV}^3 \qquad \left\langle B_s \left| \bar{s} P_L b \, \bar{s} P_R b \right| \overline{B}_s \right\rangle \simeq 0.139\,\text{GeV}^3. \qquad (35)$$

Let us recall that the naïve dimensional argument breaks down in the case of the $K$ meson where the ratio of LR over LL matrix elements has a large chiral enhancement due to the small mass of the strange quark (which amusingly kept going down in recent years).

---

[4]Note that here and below, for the LL and RR operators $\eta$ and $\mathcal{B}$ are the RG invariant correction factors (sometimes denoted $\hat{\eta}$ and $\hat{\mathcal{B}}$) while for the LR operators these are *running* factors, evaluated at $2\,\text{GeV}$ or $m_b$, respectively for $K$ and the $B$ mesons.

**Allowed New Physics contributions.** In order to discuss the allowed LR contributions, it is useful to adopt the parametrization of the allowed New Physics (NP) contributions to the K and B systems given in terms of the ratio with the SM values [26, 19, 20]:

$$h_K = \frac{\text{Re}\langle K^0|\mathcal{H}_{LR}|\overline{K}^0\rangle}{\text{Re}\langle K^0|\mathcal{H}_{SM}|\overline{K}^0\rangle}, \qquad h_\epsilon = \frac{\text{Im}\langle K^0|\mathcal{H}_{LR}|\overline{K}^0\rangle}{\text{Im}\langle K^0|\mathcal{H}_{SM}|\overline{K}^0\rangle}. \qquad (36)$$

$$h_q = \frac{\langle B_q|\mathcal{H}_{LR}|\overline{B}_q\rangle}{\langle B_q^0|\mathcal{H}_{SM}|\overline{B}_q^0\rangle}, \qquad (q=d,s) \qquad (37)$$

Notice that $h_K$ and $h_\epsilon$ are real, while $h_d$, $h_s$ are complex.

Actually, due to the uncertain long-distance parts, it is customary to normalize $h_K$ to the experimental value (and not to the contribution due to $W_L$), and to require that the LR contribution does not exceed it, regardless of the sign. We stress in fact that $\Delta M_K$ will receive a long-distance part also from the LR operators, which has so far not been estimated. We will take thus $|h_K| < 1$. As for $\epsilon_K$, we consider the limit $|h_\epsilon| < 0.3$ at 95% CL [19].

For the $B_d$ and $B_s$ mesons a slight discrepancy of the SM with data from CP-violating processes ($2.8\sigma$) was reported [19, 20]. Therefore there seems to be space for New Physics in the $B$ systems, and if the discrepancy were to be confirmed, it would even be required. We will discuss below the possibility to resolve this tension in the context of the LR model.

## 3.1 $K$: mass difference and CP violation.

The mass difference of neutral mesons is given by $\Delta m_M = 2\left|\langle M^0|\mathcal{H}^{\Delta F=2}|\overline{M}^0\rangle\right|$, and in the case of $K$, the NP contribution is dominated by the LR box diagram.

For the $K$ meson in fact, the LR contribution has a series of enhancements with respect to the LL one: on top of the factor of 8 mentioned above from the loop diagram, an important enhancement is due to the non-chiral nature of the operator: the LR matrix element is larger by a factor of $\sim 20$ with respect to the LL one, as is evident from (31). Then, another enhancement of $\sim 9$ comes from the loop function. In fact for similar left and right mixing angles, the dominant term in the sum is $i,j = c,c$, and for it we have $S_{LL} \sim x_c$ while

$$S_{LR} \simeq x_c[1 + \ln x_c + 1/4\ln(M_{W_L}/M_{W_R})] \simeq -9\, x_c. \qquad (38)$$

This turns out to be very important since the left and right mixing angles are the same in the case of $\mathcal{C}$ and turn out to be very similar in the case of $\mathcal{P}$.

In short, the LR box diagram for the $K$ mass difference has a big enhancement of $\sim 1500$ with respect to the LL one, resulting from the combined effect of: a factor 8 from the loop, the large relative factor of about 20 among the matrix elements, and the factor of 9 due to the logarithmic enhancement of the GIM, from the charm quark. For the sake of precision, let us note that the simple LL contribution in the SM falls short of a factor of $\sim 1.5$ in accounting for the experimental value, and the missing part can be ascribed to long-distance physics involving $u$-quark intermediate states. Thus when comparing the LR contribution to the experimental value, only a factor of $\sim 1000$ remains. This enhancement is the reason of the well known large bound on the LR scale in the original works [5].

The contribution of the $W_L$-$W_R$ box is expressed in terms of $h_K$ by normalizing to the experimental value $\Delta m_K^{exp}$,

$$h_K \simeq \frac{\Delta m_K^{LR}}{\Delta m_K^{exp}} = -\cos(\theta_d - \theta_s) \frac{|(V_R)_{cd}(V_R)_{cs}^*|}{|(V_L)_{cd}(V_L)_{cs}^*|} \left(\frac{2.4\,\text{TeV}}{M_{W_R}}\right)^2 \left[1 - 0.07 \ln \frac{2.4\,\text{TeV}}{M_{W_R}}\right]. \quad (39)$$

The LR contribution can have either sign depending on the phase $\theta_d - \theta_s \simeq 0, \pi$ (or the equivalent signs $S_d S_s$). Recalling that we can have at most $|h_K| \sim 1$, what we find from the $K$-meson mass difference is thus the bound $M_{W_R} \gtrsim 2.4\,\text{TeV}$. In the case of $\mathcal{P}$, the left and right mixing angles are almost equal, thus the bound. True, as shown in appendix A the relevant right angles can be somehow smaller (at most 20% at 95% CL) leading to a possible minor reduction of the bound to $2.3\,\text{TeV}$, which is of very little relevance, if any. In the case of $\mathcal{C}$ the angles are exactly equal and thus the bound of $M_{W_R} \gtrsim 2.4\,\text{TeV}$ follows directly. Actually, in view of the uncertainties in the long-distances parts, both LL and LR ones, we believe that this bound should be considered in the range $2$–$3\,\text{TeV}$.

A similar discussion applies to the operators mediated by the heavy Higgs hamiltonian (27). The process is not loop suppressed, hence the bounds on $M_H$ are higher, around $8\,\text{TeV}$.[5]

In reality both the Higgs and the $W_R$ contributions will be present at the same time (and have the same sign, since the loop function $S_{LR}$ is negative for high $M_{W_R}$, as is the Higgs exchange (27)). We have thus a correlated bound in the $M_{W_R}$–$M_H$ plane, shown in figure 1. As one can see, at low $M_{W_R} \sim 3\,\text{TeV}$ one can still have $M_H$ in the perturbative regime.

**Weak CP violation: $\epsilon_K$.** The $K$ indirect CP violation is given by the imaginary part of the same $\Delta F = 2$ hamiltonian computed above, $\epsilon_K = -\mathrm{e}^{i\pi/4} \operatorname{Im}\langle K^0 | \mathcal{H}^{\Delta F=2} | \overline{K}^0 \rangle / \sqrt{2} \Delta m_K$.[6] Let us recall that in the SM $\epsilon_K$ is reproduced by the $c$-$t$ and $t$-$t$ terms in the hamiltonian, while the much larger $c$-$c$ term, being real, does not contribute. In the LR model, because of the additional phases in $V_R$, also this term can give rise to an imaginary part, and in fact the contribution to $\epsilon_K$ is dominated by the $c$-$c$ and $c$-$t$ ones. In terms of $h_\epsilon$ we have

$$\mathcal{P}: \quad h_\epsilon \simeq \operatorname{Im}\left[\mathrm{e}^{i(\theta_d - \theta_s)}\left(A_{cc} + A_{ct}\mathrm{e}^{-i\beta}\cos(\theta_c - \theta_t)\right)\right] \quad (40)$$

$$\mathcal{C}: \quad h_\epsilon \simeq \operatorname{Im}\left[\mathrm{e}^{i(\theta_d - \theta_s)}\left(A_{cc} + A_{ct}\cos(\theta_c - \theta_t - \beta)\right)\right], \quad (41)$$

---

[5]It is worth noting the sizable $c$-$t$ contribution in the Higgs-mediated process, which can in fact cancel 40% of the total amount, for favorable phases/signs $S_c S_t = -1$. This is the choice adopted in deriving the bound. Notice also that this cancelation does not substantially happen in the LR loop, it affects only the limit on $M_H$.

[6]At the level of accuracy required here, the few percents corrections due to the phase of $K$ decay and to the deviation of the $\epsilon$ phase from $45°$ are unimportant. Note that these are not significantly altered in the LR model.

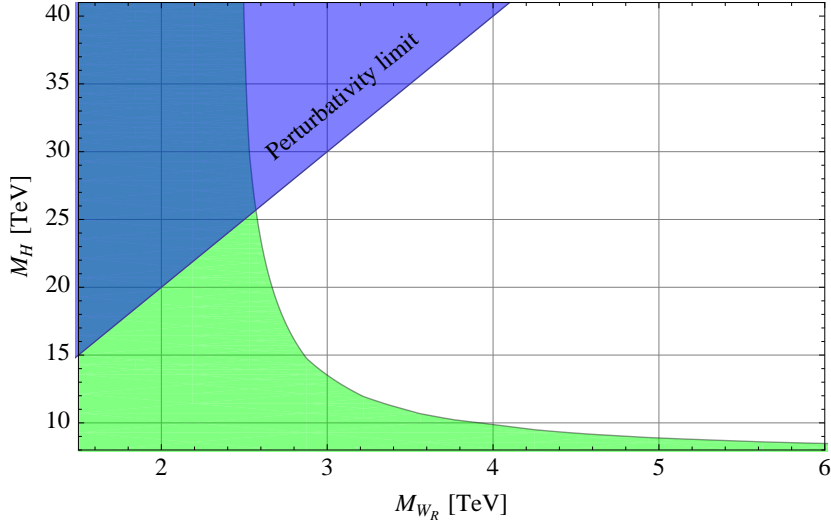

Figure 1: Correlated limit on the heavy Higgs and $W_R$ masses from $\Delta m_K$, which excludes the shaded green zone. The perturbativity limit on the Higgs mass is also shown in blue.

where $\beta = -\arg(V_{Ltd})$ and the c-c and c-t terms can be estimated as

$$A_{cc} \simeq \left[150 + 8.2 \ln\left(\frac{M_{W_R}}{2.5\,\text{TeV}}\right)\right]\left(\frac{2.5\,\text{TeV}}{M_{W_R}}\right)^2 + 84\left(\frac{15\,\text{TeV}}{M_H}\right)^2 \tag{42}$$

$$A_{ct} \simeq \left[3.8 + 1.1 \ln\left(\frac{M_{W_R}}{2.5\,\text{TeV}}\right)\right]\left(\frac{2.5\,\text{TeV}}{M_{W_R}}\right)^2 + 12\left(\frac{15\,\text{TeV}}{M_H}\right)^2. \tag{43}$$

Since we should have at most $|h_\epsilon| \lesssim 0.3$ [19], the LR contribution to $\epsilon$ can become too large for generic phases.

Note also that in addition to the enhancements discussed above for $\Delta m_K$, the c-t contribution has a further enhancement compared to the SM, by the different GIM realization: whereas in the LL diagram it is suppressed by $S_{LL} \sim x_c \propto m_c^2$, in the LR diagram it is only $S_{LR} \sim \sqrt{x_c} \propto m_c$. Thus, in case the c-c contribution is real, the c-t contribution alone, which is complex in the case of $\mathcal{P}$, would set a relatively strong bound on $M_{W_R}$. However, the (tiny) phase differences $\theta_d - \theta_s$ between $V_R$ and $V_L$ which show up in the much larger c-c term, are enough to cancel the c-t one.[7] Thus no bound effectively emerges, both for $M_{W_R}$, $M_H$ and also when both are present [12].

The case of $\mathcal{C}$ is much simpler, and one sees immediately from equation (38), that in the limit of vanishing phases there is no CP violation from the LR hamiltonian. This is a general feature of $\mathcal{C}$ as a LR symmetry. So the smallness of $\epsilon_K$ simply boils down to the requirement that the phases $\theta_d$ and $\theta_s$ are very near, $\theta_d - \theta_s \lesssim 10^{-3}$–$10^{-4}$. Possible hints toward new physics in $\epsilon_K$ (like in the interpretations recently suggested [27]) may be related to this phase difference. In any case, no bound on $M_{W_R}$ or $M_H$ results.

---

[7]In equation (37) one could have $\theta_d - \theta_s = O(10^{-2})$ or $\theta_d - \theta_s = \pi + O(10^{-2})$. The optimal scenario, in agreement with the constraints on $\Delta m_K$ and with $B$ meson mixing (see next subsection) is $\theta_d - \theta_s = O(10^{-2})$ and $\theta_c - \theta_t = \pi + O(10^{-2})$, so that $\cos(\theta_c - \theta_t) \simeq -1$. This is equivalent to choosing $S_d = S_s$ as mentioned in the previous section, and $S_c = -S_t$.

## 3.2 $B_d$ and $B_s$: mass differences and CP violation

For the $B$ mesons, the LR/LL enhancements are not present, because the chiral enhancement of the matrix element is very limited (a factor of 1.5) and in addition because the dominant loop is the one with top exchange, for which there is no large $\ln x_c$. As a result, the ratio of LR/LL contributions is of the order of 30 from which a very modest limit of $M_{W_R} \gtrsim 400\,\text{GeV}$ results.

We further observe that in addition to the LR box, also the RR one has to be taken into account, for the $B$ mesons. This is again because there is no chiral enhancement of the LR matrix element, but also because the (top-mediated) RR loop function is enhanced by the top mass (that is actually much larger than what was thought at the time of the first studies). This compensates the factor of 8 in the LR loop: $S_{RR}/8S_{LR} \sim -0.6$ and, as a result, the RR contribution is roughly one half of the LR one. It also has the opposite sign, offering a partial cancelation. Finally, because in the RR term only $V_R$ mixing matrices appear, the RR contribution leads also to a modification of CP-violation.

Collecting together the LR and RR boxes we can approximate the overall NP effect as

$$h_d \simeq 33.2\,\beta \left( 8 \frac{S_{LR}(x_t, x_t, \beta)}{S_{LL}(x_t, x_t)} r_d + \frac{S_{RR}(x_t, x_t, \beta)}{S_{LL}(x_t, x_t)} r_d^2 \right), \tag{44}$$

$$h_s \simeq 33.9\,\beta \left( 8 \frac{S_{LR}(x_t, x_t, \beta)}{S_{LL}(x_t, x_t)} r_s + \frac{S_{RR}(x_t, x_t, \beta)}{S_{LL}(x_t, x_t)} r_s^2 \right), \tag{45}$$

where $r_q = -(V_R)_{tq}(V_R)^*_{tb}/(V_L)_{tq}(V_L)^*_{tb}$.

Ignoring first their phases, we have $r_q \simeq \pm 1$, and we can approximate $h_d$ and $h_s$ as

$$|h_d| \simeq \frac{\Delta m_{B_d}^{LR}}{\Delta m_{B_d}^{exp}} \simeq \left( \frac{0.46\,\text{TeV}}{M_{W_R}} \right)^2 \left[ 1 - 0.60 \ln \frac{0.45\,\text{TeV}}{M_{W_R}} \right] \tag{46}$$

$$|h_s| \simeq \frac{\Delta m_{B_s}^{LR}}{\Delta m_{B_s}^{exp}} \simeq \left( \frac{0.47\,\text{TeV}}{M_{W_R}} \right)^2 \left[ 1 - 0.50 \ln \frac{0.47\,\text{TeV}}{M_{W_R}} \right]. \tag{47}$$

Therefore at first glance the $B$ mesons have a minor role in setting the bound on the LR scale. The allowed magnitude of NP correction to the SM has become tighter for the $B$ system, due to increased statistics in experiments and precision in the lattice evaluation of matrix elements, and currently the PDG sets a limit of 20% for NP contributions [28], which would result in the mild bounds $M_{W_R} \gtrsim 1.9\,\text{TeV}$, and $M_H \gtrsim 13\,\text{TeV}$.[8] However, this seems to be at variance with more recent global CKM analyses versus the $B$ data [19, 20], as we recalled above. The situation is also intriguing since both the $B_d$ and $B_s$ systems allow for large NP contribution, for definite nonzero ranges of CP phases. We thus discuss below together the mass difference and CP violation in the $B$-meson systems.

---
[8]These bounds are somewhat weaker than the ones reported in [12]; this is due to a seemingly discrepant factor of 2 in their $B$-mesons matrix elements, for the chosen values of $\mathcal{B}_{B_{d,s}}$, $f_{B_{d,s}}$.

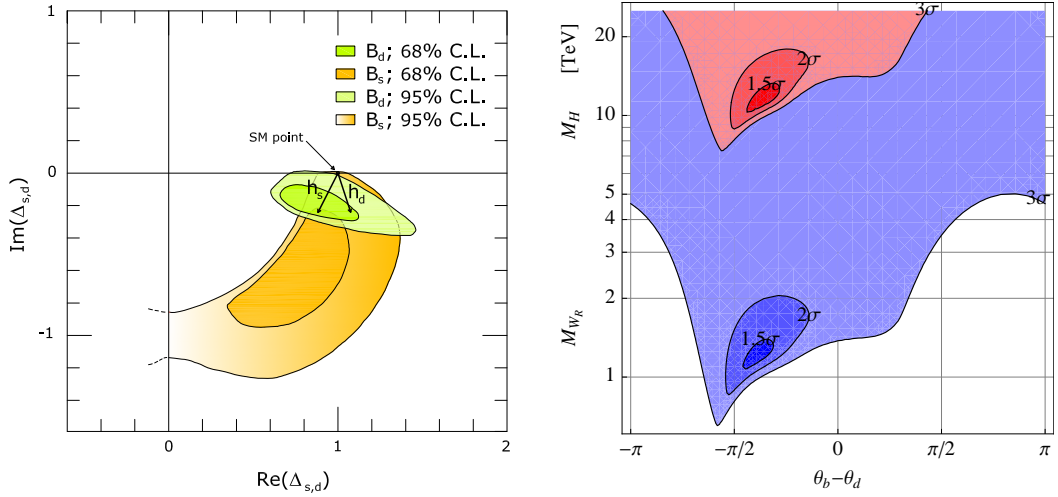

Figure 2: CP violation in the $B_d$, $B_s$ sectors in the LR model with $\mathcal{C}$. Here $\Delta_q = 1 + h_q$, for $q = d, s$, so that $h_q$ is relative to the SM point. The figure shows the allowed/preferred New-Physics contributions from unitary triangle fits (yellow and green zones) with the expected contributions in the LR model (black arrows). Right: exclusion plot for $M_{W_R}$ (lower, blue) or $M_H$ (upper, red), versus the only relevant free phase, $\theta_b - \theta_d \simeq \theta_b - \theta_s$. The $2.8\sigma$ discrepancy of the SM can be resolved for a range of $M_{W_R} = 0.5\text{--}2\,\text{TeV}$ or for $M_H = 9\text{--}18\,\text{TeV}$ (95% CL).

**Weak CP violation: the $B$ sector.** The complex ratios of New Physics to SM $h_d$, $h_s$ depend on the Right phases and mixing angles through $r_q = -(V_R)_{tq}(V_R)^*_{tb}/(V_L)_{tq}(V_L)^*_{tb}$, but in view of the similarity of L and R mixing angles, these amount almost only to phases:

$$r_d \simeq -e^{i(\theta_d - \theta_b + \phi)}, \qquad r_s \simeq -e^{i(\theta_s - \theta_b)}, \tag{48}$$

where $\phi = \arg[(V_R)_{td}/(V_L)_{td}]$. The differences between the $\mathcal{P}$ and $\mathcal{C}$ cases are entirely encoded in these quantities.

Indeed for $\mathcal{P}$ we have $\theta_{d,s,b} \simeq 0, \pi$ (the signs in equation (**??**)) and $\phi \simeq 0$, so that $r_d$, $r_s$ are both real to a good approximation. Therefore also $h_d$ and $h_s$ are real, and it is immediately clear from figure 2 that they can only worsen the tension of the SM.

For $\mathcal{C}$ instead recall that from the smallness of $\epsilon_K$ we had to fix $\theta_s \simeq \theta_d + 0, \pi$, therefore we only have one free phase, $\theta_b - \theta_d$. Also, $\phi$ is fixed to be $\phi \simeq -2\arg[(V_L)_{td}] = 2\beta \simeq 44°$, here taken at its central value. Therefore, $r_d$ and $r_s$ have correlated phases with a small separation. Then, it turns out that also $h_d$, $h_s$ point roughly in the same region, and can lead to a better agreement with present data. We stress that this is a prediction of the LR model with $\mathcal{C}$, once the constraint on $\epsilon_K$ is satisfied. As an example in the left plot in figure 2 we show a favorable configuration of $h_d$, $h_s$ pointing toward the allowed zones (using $M_{W_R} \simeq 3\,\text{TeV}$ and $M_H \simeq 15\,\text{TeV}$).[9]

Considering the $W_R$ contribution alone, it is clear that for this to happen $M_{W_R}$ can not be too high, so that in addition to a lower bound an upper bound appears. The analysis of the combined goodness of fit for $B_d$ and $B_s$ leads to the right plot in figure 2, where we show the preferred zones as a function of $M_{W_R}$ and the free phase $\theta_b - \theta_d$. We thus find $M_R = 0.7\text{–}2\,\text{TeV}$, where the LR contribution reduces the disagreement from $2.8\sigma$ to at most $2\sigma$, and a smaller region where the agreement can drop below $1.5\sigma$. A similar conclusion can be drawn regarding the contributions from the heavy FC Higgs alone (in the same plot) which results in the range 9–18 TeV.

In reality, the simultaneous contribution of $W_R$ and $H$ has to be considered, and this leads to a correlated bound on their masses, that we display in figure 3. From this analysis we find that the region where the discrepancy is resolved extends down to scales reachable at LHC. Also, the heavy Higgs is bound to be mostly in the perturbative regime, since we already know from $K$-oscillations that $M_{W_R} > 2.5\,\text{TeV}$. In this respect, we observe that while the limit of infinitely heavy Higgs $H$ is not admissible because of perturbativity, at the other end one may decouple $W_R$ and keep $H$ at low scale ($M_H \sim 9\text{–}20\,\text{TeV}$). However, this goes in the direction of a 2HDM rather than a Left-Right symmetric model. As a result, the most interesting zone is for low $W_R$ and not too high $H$. In this region the $B$-mesons CP-violation discrepancies are resolved nicely and $W_R$ is in the LHC range. Recently, another tension with the SM emerged from the D0 data on dimuon semileptonic anomaly at the level of $3\sigma$ [29]. It is worth investigating whether the minimal LR theory can account for it.

In conclusion, the $B$-mesons system do not lead to any new limits on the $W_R$ or heavy Higgs scales. On the contrary their presence can play a positive role in explaining the present mild discrepancy of the SM with data.

### 3.3 Weak CP violation: $\epsilon'$.

The 'direct' CP violation in $\Delta S = 1$ processes is measured by

$$\epsilon' = \frac{i}{\sqrt{2}} \omega \left( \frac{\text{Im} A_2}{\text{Re} A_2} - \frac{\text{Im} A_0}{\text{Re} A_0} \right) \frac{q}{p} e^{i(\delta_2 - \delta_0)}, \qquad (49)$$

where $p$, $q$ are the $K^0$, $\overline{K}^0$ mixing parameters, $A_0$, $A_2$ are the decay amplitudes of $K \to 2\pi$ in $I = 0$, $2$ isospin states, and $\omega \equiv A_2/A_0$. The ratio $p/q \simeq 1$ with an excellent approximation. The amplitudes $A_I$ are defined from the $\Delta S = 1$ effective Hamiltonian as $\langle (2\pi)_I | (-I) H_{\Delta S=1} | K^0 \rangle = A_I e^{i\delta_I}$, where $\delta_I$ are the strong phase of $\pi\pi$ scattering.

---

[9]The smallness of $h_s$, $h_d$ ensures that the better known ratio $\Delta m_{B_d}/\Delta m_{B_s}$ is also not spoiled.

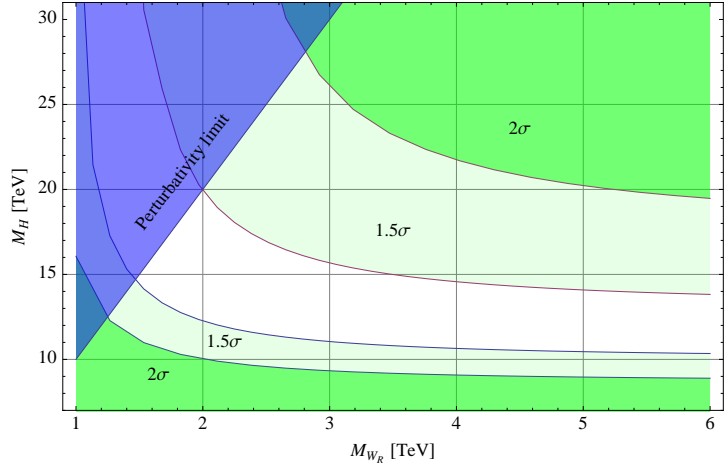

Figure 3: $\Delta m$ and CP violation for $B_d$, $B_s$, and the LR model with $\mathcal{C}$. The simultaneous contribution of $W_R$ and $H$ leads to the preferred region shown in white in the $M_{W_R}$–$M_H$ plane.

In the SM $\epsilon'$ is due to both QCD and QED penguin diagrams [30] which have opposite signs so that the prediction for the net effect is plagued by uncertainties in the relevant hadronic matrix elements. One can still use this parameter to obtain a bound on the right-handed scale, keeping in mind that a discrete cancelation with the SM might take place.

In the LRSM there are tree-level contributions to the $\Delta S = 1$ amplitude, and CP violation is given by phases in $V_R$ as well as by the phase $\alpha$ in the $W_L$-$W_R$ mixing, $\tan\zeta\, e^{i\alpha} = e^{i\alpha}(M_{W_L}^2/M_{W_R}^2)2x/(1+x^2)$. The Hamiltonian due to the tree-level exchange of left and right gauge bosons can be written as [9, 31][10]

$$H_{\Delta S=1} = \sqrt{2} G_F \lambda_u^{LL} \left[ \left( \frac{\alpha_S(\mu^2)}{\alpha_S(M_{W_L}^2)} \right)^{-\frac{2}{b}} O_+^{LL}(\mu) + \left( \frac{\alpha_S(\mu^2)}{\alpha_S(M_{W_L}^2)} \right)^{\frac{4}{b}} O_-^{LL}(\mu) \right] \qquad (50)$$

$$+ \sqrt{2} G_F \frac{M_{W_L}^2}{M_{W_R}^2} \lambda_u^{RR} \left[ \left( \frac{\alpha_S(\mu^2)}{\alpha_S(M_{W_R}^2)} \right)^{-\frac{2}{b}} O_+^{RR}(\mu) + \left( \frac{\alpha_S(\mu^2)}{\alpha_S(M_{W_R}^2)} \right)^{\frac{4}{b}} O_-^{RR}(\mu) \right]$$

$$+ 2\sqrt{2} G_F \sin\zeta\, \lambda_u^{LR} e^{i\alpha} \left[ \left( \frac{\alpha_S(\mu^2)}{\alpha_S(M_{W_L}^2)} \right)^{\frac{8}{b}} O_-^{LR}(\mu) + \left( \frac{\alpha_S(\mu^2)}{\alpha_S(M_{W_L}^2)} \right)^{-\frac{1}{b}} O_+^{LR}(\mu) \right]$$

$$+ 2\sqrt{2} G_F \sin\zeta\, \lambda_u^{RL} e^{-i\alpha} \left[ \left( \frac{\alpha_S(\mu^2)}{\alpha_S(M_{W_L}^2)} \right)^{\frac{8}{b}} O_-^{RL}(\mu) + \left( \frac{\alpha_S(\mu^2)}{\alpha_S(M_{W_L}^2)} \right)^{-\frac{1}{b}} O_+^{RL}(\mu) \right].$$

---

[10]The tree-level contribution to $\epsilon'$ from the heavy Higgs are negligible [9] being suppressed by two Yukawa entries and the large $M_H$ scale. Also, penguins involving $W_R$ are suppressed by the loop factor.

Here as before, $\lambda_u^{AB} = V_{ud}^{A*}V_{us}^B$. The renormalization through each intermediate scale has $b = 11 - 2N_f/3$ with $N_f$ active quark flavors. The four quark operators are

$$O_{\pm}^{LL,RR} = \bar{d}\gamma^\mu P_{L,R} u \bar{u}\gamma_\mu P_{L,R} s \pm \bar{d}\gamma^\mu P_{L,R} s \bar{u}\gamma_\mu P_{L,R} u$$

$$O_{+}^{LR,RL} = \bar{d}\gamma^\mu P_{L,R} u \bar{u}\gamma_\mu P_{R,L} s + \frac{2}{3}\bar{d}\gamma^\mu P_{R,L} s \bar{u}\gamma_\mu P_{L,R} u$$

$$O_{-}^{LR,RL} = \frac{2}{3}\bar{d}\gamma^\mu P_{R,L} s \bar{u}\gamma_\mu P_{L,R} u \tag{51}$$

whose matrix elements are estimated at $\mu = 1\,\text{GeV}$ as [31],

$$\langle (2\pi)_{I=0}|O_+^{LL,RR}|\overline{K}^0\rangle = \pm\frac{X}{3\sqrt{3}}, \qquad \langle (2\pi)_{I=0}|O_+^{LR,RL}|\overline{K}^0\rangle = \pm\frac{4X}{9\sqrt{3}}$$

$$\langle (2\pi)_{I=2}|O_+^{LL,RR}|\overline{K}^0\rangle = \pm\frac{2\sqrt{2}X}{3\sqrt{3}}, \qquad \langle (2\pi)_{I=2}|O_+^{LR,RL}|\overline{K}^0\rangle = \pm\frac{2\sqrt{2}X}{9\sqrt{3}}$$

$$\langle (2\pi)_{I=0}|O_-^{LL,RR}|\overline{K}^0\rangle = \pm\frac{X}{2\sqrt{3}}, \qquad \langle (2\pi)_{I=0}|O_-^{LR,RL}|\overline{K}^0\rangle = \mp\frac{1}{\sqrt{3}}\left(\frac{X}{18} + \frac{Y}{2} + \frac{Z}{6}\right)$$

$$\langle (2\pi)_{I=2}|O_-^{LL,RR}|\overline{K}^0\rangle = 0, \qquad \langle (2\pi)_{I=2}|O_-^{LR,RL}|\overline{K}^0\rangle = \mp\frac{\sqrt{2}}{6\sqrt{3}}\left(\frac{X}{6} - Z\right), \tag{52}$$

where $X$, $Y$ and $Z$ are

$$X \equiv -\langle\pi^-|\bar{d}\gamma_\mu\gamma_5 u|0\rangle\langle\pi^+|\bar{u}\gamma^\mu s|\overline{K}^0\rangle = i\sqrt{2}f_\pi(m_K^2 - m_\pi^2) \simeq 0.03i\,\text{GeV}^3$$

$$Y \equiv -\langle\pi^+\pi^-|\bar{u}u|0\rangle\langle 0|\bar{d}\gamma_5 s|\overline{K}^0\rangle = i\sqrt{2}f_K A^2\left(1 - \frac{m_K^2}{m_\sigma^2}\right)^{-2} \simeq 0.273i\,\text{GeV}^3$$

$$Z \equiv -\langle\pi^-|\bar{d}\gamma_5 u|0\rangle\langle\pi^+|\bar{u}s|\overline{K}^0\rangle = i\sqrt{2}f_\pi A^2\left(1 - \frac{m_\pi^2}{m_\sigma^2}\right)^{-2} \simeq 0.18i\,\text{GeV}^3 \tag{53}$$

with $A \equiv m_K^2/(m_s + m_d)$, and $f_{\pi,K}$ the decay constants for $\pi$ and $K$.

It is again useful to discuss separately the bounds on $M_{W_R}$ in the cases of $\mathcal{P}$ and $\mathcal{C}$.

**Case of $\mathcal{P}$**  From (47) it is clear that $H_{\Delta S=1}$ is a function of $M_R$, $\alpha$ and $x = v_1/v_2$. In [12] the values of $\alpha \sim 0.1$ and $x \sim m_b/(2m_t)$ were found to satisfy the $n$EDM constraints, so that $H_{\Delta S=1}$ is a function of $M_R$ only. Using the experimental values of $\text{Re}A_0$ and $\text{Re}A_2$ in (46) it is therefore possible to plot $\epsilon'$ as function of $M_{W_R}$. As in [12], assuming all $S_q = 1$, and require that $\epsilon'$ does not exceed the experimental value, we find $M_{W_R} > 4.2\,\text{TeV}$. On the other hand, for different choice of signs, in particular setting $S_u = -1$, which is allowed once one ignores the $n$EDM constraints (see next section) it is possible to reduce the lower limit to $M_{W_R} > 3.1\,\text{TeV}$.

**Case of $\mathcal{C}$**  As before with $\epsilon$, the free phases can be chosen as to make $\epsilon'$ vanish. In particular, the RR contribution is small because $\theta_d$ and $\theta_s$ have to be near. The remaining two contributions have the oposite sign and their sum is proportional to $\sin(\alpha + \theta_u)$. As a result, no bound on $W_R$, $H$ masses emerges.

## 3.4 EDM and the strong CP problem

At first glance, due to the required smallness of the strong CP phase $\bar{\theta}$, one would think that this could be the source of the best limit on the LR scale. The situation is rather subtle though, and in general no bound comes out, because one can always fine tune. As we all know, the electric dipole moment of the neutron sets an upper bound $\bar{\theta} \lesssim 10^{-10}$ and this normally implies a strong bound on CP-violating phases. The crucial point is that this does not apply to the KM phase $\delta$ of the SM since the perturbative contribution to the $\bar{\theta}$ appears only at the three-loop level and is negligibly small [32, 33]. There is an important question then, as to why the original strong CP parameter $\theta$ is so small, but that is not a problem: once chosen small it stays small in the SM. In other words, in the SM weak CP plays effectively no role for the strong CP.

The situation is completely different for the other phases, for they enter directly in the strong CP phase at the tree level. The nice features of the small perturbative contribution to $\bar{\theta}$ from the weak phase is lost in these models as typical of beyond the SM physics. One is forced to fine tune the resulting phase against the original QCD angle. This may not be appealing but is phenomenologically correct, before one has a theory of the strong CP phase. In short, one can not put limits on the LR scale from the EDM, unless one forbids fine tuning. This however should be considered as a theoretical rather than a phenomenological limit, and it becomes a matter of taste whether one sticks to it or not.

Here we disagree with [12, 13] in the case of $\mathcal{P}$. After using $\alpha$ to make $\epsilon$ small, they argue that EDM then forces $W_R$ to be heavier than about 10 TeV. The trouble is that the large $\alpha$ enters into the overall $\bar{\theta}$ which then has to be fine tuned anyway. Clearly one can and should perform the fine tuning at the end of the day as to make the total contribution to the EDM small. In other words, both $\epsilon$ and EDM can be made small by a judicious choice of $\alpha$ and $\theta$. What remains is the limit from $\epsilon'$ which can not be rotated away.

The question is what happens if one refuses to fine tune the physically separate contributions from the strong and weak sectors. In that case, one must opt for hermitian quark matrices, or manifest LR symmetry, i.e. vanishing $\alpha$ (or at least smaller than $10^{-10}$). In that case, it is $\epsilon$ that sets a huge limit on $W_R$ mass, $M_{W_R} > 15$ TeV [12].

In the case of $\mathcal{C}$, one does not have that option and one must fine tune the strong CP parameter. Notice that this can be done by exploiting the "up" phases, without spoiling the weak-CP-violating observables, where only the "down" phases enter. The precise amount of fine-tuning is worth investigating, but is beyond the scope of this work.

## 3.5 Summary of bounds

We summarize here the limits from both CP-conserving and CP-violating observables, separately in the two cases of LR symmetry, $\mathcal{P}$ or $\mathcal{C}$.

**Case of P**

|             | $M_{W_R}$[TeV] | $M_H$[TeV] |
|-------------|:--------------:|:----------:|
| $\Delta m_K$      | 2.3    | 7.7   |
| $\Delta m_{B_d}$  | 1.9    | 13    |
| $\Delta m_{B_s}$  | 1.8    | 12    |
| $CP$ in $B_{d,s}$ | ×      | ×     |
| $\epsilon_K$      | −      | −     |
| $\epsilon'$       | 3.1 (4)| −     |
| $n$EDM            | − (8)  | − (25)|

The limits from $n$EDM in parentheses are not really valid since they assume $\bar{\theta} = 0$. The cross × refers to the impossibility to resolve the current $3\sigma$ tension of the SM with data in $B_{d,s}$ CP violation.

**Case of C**

|                              | $M_{W_R}$[TeV] | $M_H$[TeV] |
|------------------------------|:--------------:|:----------:|
| $\Delta m_K$                 | 2.5            | 7.7        |
| $\Delta m$ and CP in $B_{d,s}$ | 0.5          | 6          |
| $\epsilon_K$, $\epsilon'$, $n$EDM | −        | −          |
| Best point                   | 2.8            | 15         |

The best point in the last row refers to a good scenario where the LR symmetry as $\mathcal{C}$ can satisfy all bounds described above, can resolve the present tension of the SM with $B$ CP violation, and is detectable at LHC.

## 4 Experimental Signatures

It is good to recall the main motivations for studying LR symmetry. The main of course is the LR symmetry itself whose origin has attracted physicists ever since the discovery of parity violation. An other important motivation is the naturalness of the seesaw mechanism [34] which in this theory takes a particularly appealing form: the smallness of neutrino masses is linked to the breakdown of LR symmetry at the scale $M_R \gg M_W$. If $M_R$ is to be accessible at the colliders, then the Yukawa Dirac couplings must be small, which is of course a natural possibility, protected by chiral symmetry. Furthermore, small Yukawa couplings are already one of the most striking features of the SM. Still, it is always possible to imagine large Dirac Yukawa couplings which then miraculously cancel in order to provide small neutrino masses. We will clearly not adhere to this possibility that destroys the beauty of the seesaw mechanism and one of the main motivations for this work. We thus stick to the seesaw scenario, with small Yukawa Dirac couplings and RH neutrino masses relevant for LHC.

We describe below the experimental signatures that are expected in this scenario, starting from the most compelling case of direct detection of the $W_R$ at LHC via Lepton Number

Violation (LNV). We then briefly describe the possible contributions of the LR model to neutrinoless double beta decay ($0\nu\beta\beta$) and Lepton Flavour Violation (LFV).

## 4.1 Collider

The low-scale $W_R$ will be effectively produced at LHC through Drell-Yan process as in figure 4.

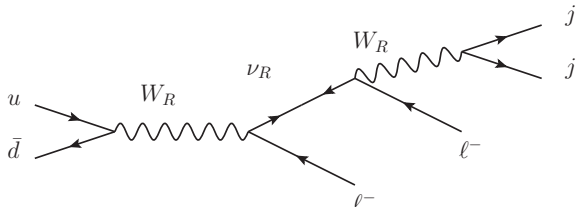

Figure 4: $W_R$ production and decay through $\nu_R$.

Once the right-handed gauge boson is produced, it will decay into jets or into a right-handed neutrino and a charged lepton. The right-handed neutrino in turn decays again through the $W_R$, and being a Majorana particle, decays equally often into a charged lepton or anti-lepton, plus jets. One has then the exciting events of same-sign lepton pairs and two jets, as a clear signature of lepton number violation. This is a collider analog of neutrino-less double beta decay, and it offers

a) the direct test of LR parity restoration through a discovery of $W_R$,

b) the direct test of lepton number violation through a Majorana nature of $\nu_R$,

c) the determination of $W_R$ and $\nu_R$ (invariant) masses.

*LHC reach.* The lepton-number violating character of the process makes it the golden channel for the discovery of $W_R$ and $\nu_R$, as envisaged in [16]. Indeed the background for this process, mainly coming from $t\bar{t}$ events, is practically absent beyond the TeV energy when one selects same sign dileptons. At the same time the invariant mass of the hardest jets plus one or two leptons, allows for a clean determination of the $\nu_R$ and $W_R$ masses. See figure 5.

Detailed studies conclude an easy probe of $W_R$ up to 4(2) TeV and $\nu_R$ in 100 MeV–2(1) TeV for a collision energy of 14(7) TeV and an integrated luminosity of 30 fb$^{-1}$ at Atlas [35] and CMS [36]. This is interesting since as we showed above, by defining LR symmetry as $\mathcal{C}$ the freedom in the additional phases leaves only the CP-conserving bound from $\Delta m_K$, effectively around $M_{W_R} \simeq 2.5$ TeV.

The LHC reach is obviously affected by the beam energy, but due to the lepton-number-violation, it is not much affected by the integrated luminosity, at least for lower scale $M_{W_R}$. In fact in the lucky scenario of $M_{W_R} \simeq 2.5$ TeV and $M_{\nu_R} \simeq 0.5$ TeV the $W_R$ could well be discovered as soon as after a few months of running at low luminosity, i.e. 200–300 pb$^{-1}$ [36, 37].

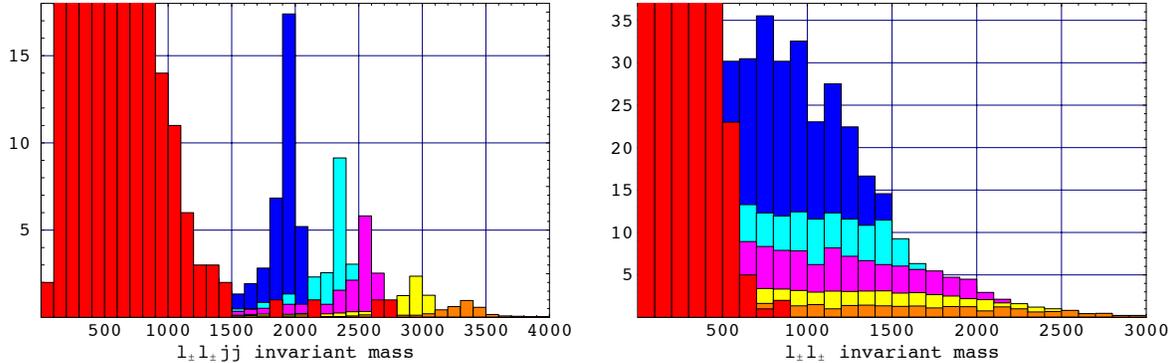

Figure 5: Selected events for $\ell^\pm\ell^\pm jj$ (left) and $\ell^\pm\ell^\pm$ (right) as a function of the invariant mass (100 GeV bins) for L = 8 fb$^{-1}$ and $E_{c.o.m.} = 14$ TeV. The $t\bar{t}$ background is the foremost (red) distribution. Here $M_{W_R}$ [TeV] is taken to be: 2.0; 2.4; 2.6; 3.0; 3.4 and $M_{\nu_R} = 1$ TeV. In the left plot the invariant mass of the $W_R$ is reconstructed nicely and the process has virtually zero background above 1.5 TeV. A similar plot, of the invariant mass of the second-hardest lepton plus the two jets, yields the invariant mass of $\nu_R$. In the right plot the invariant mass of same-sign dileptons can give a high event count with low background, before the actual peak discovery (Events generated with Pythia 6.4, Tauola; PGS detector; standard $t\bar{t}$ cuts).

In this respect it is also interesting to note that by considering just the dileptons (same sign or not) although one lacks the $W_R$ peak in their invariant mass, the signal/background ratio is favorable even for very low luminosities. This is evident from the right plot of figure 5. The dilepton invariant mass is thus efficient as an early search strategy.

*Displaced Vertex.* The scenario of low RH neutrino, with mass in the range $M_{\nu_R} \sim 5$–$100$ GeV is also interesting, since the RH neutrino lifetime becomes longer, possibly leading to a displaced vertex. For instance $\tau_{\nu_R} \gtrsim 1.5$ cm for $M_{\nu_R} \lesssim 10$ GeV (assuming $M_{W_R} = 3$ TeV). One would thus have a quite unmistakable signal of a displaced vertex with two (typically merged) jets and a nearby lepton, same sign with an other (circa opposite) lepton. A detailed study of this signature, beyond the scope of this work, is still missing. We point out that this scenario of low $\nu_R$ implies very small Yukawa couplings to have realistic neutrino masses (and barring cancelations as mentioned above) and thus the Yukawa decay channel is much more suppressed.

*Flavor structure.* The process just described of Drell-Yan $W_R$ production and $\nu_R$ decay is free from Yukawa couplings, and directly probes the Majorana mass of RH neutrinos. The observation of the RH neutrino peak and possibly of different peaks related to the three $\nu_R$ mass eigenstates, together with the different flavor channels $ee$, $\mu\mu$, $e\mu$, etc., will thus allow a direct measure of the RH (and LH) neutrino Majorana mass matrix. This may then be put into correspondence with Lepton Flavor Violations (see later). A realistic analysis of the detection possibilities in the different channels would require an update of the Montecarlo event generators, which currently include only a flavor-universal implementation of the LR model.

**Other processes.** In addition to the most promising striking signature of lepton violation via the $W_R$ Drell-Yan production described above, other processes characteristic of the LR model are possible, involving the doubly charged higgses $\Delta_{L,R}^{\pm\pm}$, or the Dirac neutrino mass matrices. They have been studied extensively in the literature and we describe them briefly.

*Ordinary Drell-Yan.* It is worth noting that the same signatures can be studied in the SM with $\nu_R$ produced by an ordinary $W_L$ [38], but its observation requires large Dirac Yukawa couplings and their miraculous cancellations in order to keep neutrino masses small. When a protection symmetry is called for, one ends up effectively with lepton number conservation and the phenomenon disappears [39].

*Doubly charged Higgses.* The $L-R$ theory possesses naturally also type II seesaw [17]. The type II offers another potentially interesting signature: pair production of doubly charged Higgses which decay into same sign lepton (anti lepton) pairs [40]. This can serve as a determination of the neutrino mass matrix in the case when type I is not present or very small [41], but in any case directly probes the flavour structure of the neutrino Majorana mass matrices.[11] One possibility in this direction is the Drell-Yan production of $Z/Z_R$, decaying into a couple of opposite-sign $\Delta^{\pm\pm}\Delta^{\mp\mp}$, which in turn decay into *two pairs* of same-sign leptons. Each couple of same-sign leptons has a wide separation (almost back-to-back) and the event is thus very striking. The double production of $\Delta$ limits the probe to at most $m_\Delta \sim 0.8$-$1\,\text{TeV}$ [42, 43, 41, 40, 44] ($m_\Delta \sim 1$-$1.5\,\text{TeV}$ only for low lying $Z_R$) but if observed these events can probe the $Z_R$-mass quite cleanly (in addition to the standard channels like dileptons with a high invariant mass). An other possibility is production of a single $\Delta_{L,R}^{\pm\pm}$ via the fusion of either $W_L$ or $W_R$ [42]. The $\Delta_{L,R}^{\pm\pm}$ then decays again into a clean same-sign dilepton. While the production of $\Delta_L$ is suppressed by its small VEV $\langle\Delta_L\rangle$, on the other hand the large VEV $\langle\Delta_R\rangle \sim v_R$ makes the production of $\Delta_R^{\pm\pm}$ interesting, even though as any fusion processes the available energy is limited to $\sim 1\,\text{TeV}$. If observed, this process would directly give information on the neutrino Majorana mass matrix.

## 4.2 Neutrinoless double beta decay

The neutrinoless double beta decay ($0\nu\beta\beta$) receives new contributions in the LR model due to the exchange of one or two $W_R$, in place of $W_L$, and there are thus three contributions which can be called LL, LR, RR. In the seesaw picture that we are pursuing, with generic Yukawa Dirac couplings (no cancelations) the LR contribution is easily shown to be negligible compared to the RR one. Whereas the LL contribution is proportional to the light neutrino mass matrix, the RR one is proportional to the inverse of the heavy neutrino mass matrix (as long as they are heavier than the exchange momentum around $\sim 100\,\text{MeV}$)

$$LL \propto [M_\nu]_{ee}, \qquad RR \propto \left[M_{\nu_R}^{-1}\right]_{ee} \qquad (54)$$

---

[11]It is worth commenting that the minimal supersymmetric left-right symmetric model [45] predicts doubly charged scalars at the collider energies [45, 46, 47] even for large scale of left-right symmetry breaking.

A priori there is no connection between the two mass matrices, so one can not make any predictions for RR. The only hope is the probe of the flavour structure of $M_{\nu_R}$ at LHC, along the lines discussed above. This requires some 'luck' since all three heavy neutrinos should be in the accessible range, and all the flavour channels should be observable.

It is noteworthy that the scenario of low LR scale to be observable at LHC considered here is consistent with observable RR contributions to $0\nu\beta\beta$ decay. It can in fact be shown that for the RH scale in the TeV region, the RR contribution is as large as the LL one generated by $M_\nu$ of the order of $1\,\text{eV}$. More precisely,

$$\left[\left(\frac{M_{\nu_R}}{\text{TeV}}\right)^{-1}\right]_{ee} \left(\frac{\text{TeV}}{M_{W_R}}\right)^4 \longleftrightarrow \left[\frac{M_\nu}{0.4\,\text{eV}}\right]_{ee}. \qquad (55)$$

For instance, a positive evidence in the current probes for $0\nu\beta\beta$ below the eV could be accounted by a contribution from the LR model with e.g. $M_{W_R} \simeq 3\,\text{TeV}$ and RH neutrino mass of the order of $10\,\text{GeV}$ (ignoring the mixings). This could lead to a displaced vertex at LHC as discussed above.

Also the exchange of doubly charged Higgs fields $\Delta_{L,R}^{--}$ gives rise to $0\nu\beta\beta$ processes. Assuming again no cancellations, the process mediated by $\Delta_L$ is negligible, since it is suppressed as $p^2/M_\Delta^2$ with respect to the LL one. The $\Delta_R$-induced contribution depends clearly on the mass ratio $M_{\nu_R}$ divided by $M_\Delta^2$, which on the other hand may even dominate over all other contributions. A detailed analysis is necessary and is now underway [48].

## 4.3 Lepton flavor violation

There is a number of sources of lepton flavour violation (LFV) in the LR theory, through the exchanges of $W_R$ and $\nu_R$ and $\Delta_L$, $\Delta_R$.

The most sensitive process appears to be $\mu \to 3e$ which is induced at tree level through the doubly charged $\Delta_{L,R}^{--}$. This is controlled by its Yukawa coupling to fermions which in turn is proportional to the heavy neutrino mass matrix, $y_\Delta = gM_{\nu_R}/M_{W_R}$. Again, the observation at LHC of its flavour structure would allow for the test of the theory, in case the $\Delta$'s are light enough.

An important role in restricting the parameter space of the theory is played by the $\mu \to e\gamma$ process and the $\mu \to e$ conversion in nuclei, whose sensitivity is planned to be improved by four to six orders of magnitude [49, 50]. In this case one could even probe leptonic CP violation, as argued in [52]. All this requires a careful quantitative analysis [48] along the lines of [53].

## 5 Conclusions and Outlook

LR symmetric theories have been for some decades one of the principal beyond the SM theories. They cure in a very elegant manner the chiral asymmetry of the standard model, and as such offer a possibility of seeing directly the restoration of parity. They also lead naturally to the seesaw-mechanism, which has become a paradigm for understanding the smallness of neutrino masses. If the scale of LR symmetry is low enough one would also see directly the violation of lepton number in colliders and thus probe the Majorana nature of neutrino mass. The exchange of right-handed neutrinos and gauge bosons can also dominate neutrino-less double-beta decay, an other important lepton-number violating process. For these reasons, the importance of the limit on the LR scale can not be overemphasized.

While in generic models this scale can be as low as 300 GeV, in the minimal theories there is a strong limit from $K$-meson physics. We revisited this important issues, studied at lengths for almost three decades. In this work we provided for the first time a complete study of the left and right mixing angles, essential for deriving a real bound on the LR scale. The essence of what we have learned is that the LR scale is allowed to be as small as 2–3 TeV, as long as charge conjugation plays the role of a fundamental LR symmetry imposed on the Lagrangian. In this case, there is no limit from CP-violating observables, and one must only make sure that the neutral meson mass differences are in accord with experiment. The CP violation can potentially lead to somewhat more stringent bounds. This affects the case of $\mathcal{P}$ as a LR symmetry, where one is hit by the bounds from the CP violating parameters $\epsilon$ and/or $\epsilon'$. One can fine-tune one of them and ideally one should do it for $\epsilon$ since $\epsilon'$ is subject to considerable uncertainties. This is what is done by [12] which finally obtains a large limit of $M_{W_R} \gtrsim 4$ TeV.

There is also $n$EDM to be taken into consideration. However this is obscured by the strong CP parameter $\bar{\theta}$ and therefore no phenomenological limits can be put on this ground. If one refuses such a fine tuning one can obtain a huge limit on the LR scale of the order of 10 TeV which would make it completely outside the LHC reach. Alternatively, one can use the phase freedom to fine tune separately $\bar{\theta}$ and the weak contribution to $n$EDM can be made small by constraining the CP phases. This is what is done in reference [12], which then results in the above quoted limit from $\epsilon'$. Avoiding this procedure and fine tuning the overall $n$EDM one can lower this limit to $M_{W_R} \gtrsim 3$ TeV. In other words, on phenomenological grounds, even in the case of $\mathcal{P}$ one can have LR symmetry accessible at LHC. This is one of our central results. What about $n$EDM in the case of $\mathcal{C}$? As we discussed above, one has sufficiently many phases to make it small, without setting any limit on the LR scale.

We analyzed carefully also the bounds from oscillations and CP violation in the $B_d$, $B_s$ sector. Here, the stringent limits and the current mild ($3\sigma$) discrepancy of the SM with data poses a tough challenge to the models. While the case of $\mathcal{P}$ can not generate the right CP violation, we find that in the case of $\mathcal{C}$ not only there are enough phases to account for the observed CP violation, but the model predicts also the right correlated phases of $B_d$ and $B_s$. This works nicely for $W_R$ accessible at LHC and the heavy $H$ in the perturbative regime.

In summary, charge conjugation, an appealing gauged LR symmetry, automatically present in the SO(10) grand unified theory, allows the right-handed charged gauge boson $W_R$ to weigh as little as about 2–3 TeV when all the uncertainties are taken account, corresponding to about 3.5–5 TeV for the new neutral gauge boson $Z_R$. This could even offer an opportunity for the discovery of both charged and neutral gauge bosons at LHC, with their spectacular signatures of direct lepton-number violation. Although the theory can not predict this scale, we believe that in view of the importance of possible parity restoration and lepton number violation, the search for LR symmetry should be one of the priorities at LHC.

**Acknowledgements.** We thank A. Melfo and F. Vissani for collaboration in the initial stage of this work, and V. Tello and Y. Zheng for discussion. The work of M.N. was supported by the Slovenian Research Agency and by the Deutsche Forschungsgemeinschaft via the Junior Research Group SUSY Phenomenology within the Collaborative Research Centre 676 Particles, Strings and the Early Universe. The work of G.S. was partially supported by the EU FP6 Marie Curie Research and Training Network "UniverseNet" (MRTN-CT-2006-035863).

———

## A   Quark mixing matrices: Case of $\mathcal{P}$

As shown in (17), in the case of $\mathcal{P}$ as LR symmetry, the Yukawa couplings are hermitian, and thus the hermiticity of quark mass matrices depends crucially on the complexity of the bi-doublet vevs. If the vevs are real, the matrices are hermitian therefore left and right mixing angles are exactly the same. This is normally called 'manifest LR symmetry'. Our study below shows that the LR symmetry in the $\mathcal{P}$ case ends up being quite manifest: the deviations are rather small.

This follows from the result that even for large $x \sim O(1)$, the 'spontaneous' phase $\alpha$ is small, $e^{i\alpha} \simeq -1$. Therefore either because of $x \ll 1$ or $e^{i\alpha} \simeq -1$, mass matrices are approximately hermitean. This can be understood analytically in the general case, i.e. without assuming small $x = v_2/v_1$, by applying the Gershgorin theorem on complex matrices, to the mass matrix of 'up' quarks.

**The spontaneous phase.**   We start by redefining equation (9) by multiplying $M_d$ by $e^{-i\alpha}$, and for simplicity we introduce $Y_{u,d} = M_{u,d}/v$:

$$Y_u = c\, Y + s\, e^{-i\alpha}\, \tilde{Y}, \qquad Y_d = s\, Y + c\, e^{-i\alpha}\, \tilde{Y} \tag{56}$$

We choose to work in a basis where $Y$ is diagonal, $Y = \mathrm{diag}\{Y_1, Y_2, Y_3\}$, and we can take $Y_3 > 0$ (by an overall sign) and also $\tilde{Y}_{33} > 0$ (by shifting $\alpha$). Of course, all $Y_i$, $\tilde{Y}_{ii}$ are real. Two of the off-diagonal elements of $\tilde{Y}$ can be chosen real by a phase multiplication, which does not affect the mixing matrices.

Next, from the unitary bidiagonalization of both $Y_u$ or $Y_d$, we know that all elements are smaller than the sum of the eigenvalues, i.e.

$$|Y_{u\,ij}| \leq \lambda_t + \lambda_c + \lambda_u \simeq \lambda_t \tag{57}$$

$$|Y_{d\,ij}| \leq \lambda_b + \lambda_s + \lambda_d \simeq \lambda_b, \tag{58}$$

where $\lambda_{u,c,t}$ and $\lambda_{d,s,b}$ denote the eigenvalues of $Y_u$ and $Y_d$. These conditions become:

$$c|\tilde{Y}_{ij}| \leq \lambda_b \quad \text{for } i \neq j, \tag{59}$$

$$|sY_3 + c\tilde{Y}_{33}e^{-i\alpha}| \leq \lambda_b \quad \text{for } i = j = 3, \tag{60}$$

$$|cY_3 + s\tilde{Y}_{33}e^{-i\alpha}| \leq \lambda_t \quad \text{for } i = j = 3. \tag{61}$$

Since the off-diagonal elements of $Y_u$ and $Y_d$ are related, the first of these equations implies that also the off-diagonal elements of $Y_u$ are small, in particular for $i \neq j$ we have $Y_{u\,ij} < x\lambda_b < \lambda_b$.

Now, we can assume $Y_{u33} > Y_{u22}, Y_{u11}$ (this is possible without loss of generality, by rearranging $Y_{u\,ii}$ with a transformation of both matrices $Y_{u,d}$) and we see that $Y_{u\,33}$ has to be close to $\lambda_t$. In turn, $Y_{u11}$ and $Y_{u22}$ have to be smaller than $\lambda_b$, otherwise 'charm' and 'up' would end up being too heavy.

Now we apply the Gershgorin theorem [54] to the matrix $Y_u$. This theorem states that the eigenvalues of any complex matrix (here 3×3) lie, in the complex plane, inside the union of the three Gershgorin circles; these are built by taking the diagonal elements as their centers and the absolute value of the sum of nondiagonal elements in the relative column as their radii. In our case, according to the previous discussion, we have for the centers $C_{u\,i}$ and the radii $R_{u\,i}$:

$$C_{u\,1} = Y_{u11} < \lambda_b, \qquad R_{u\,1} = s(|\tilde{Y}_{21} + \tilde{Y}_{31}|) < 2x\lambda_b < 2\lambda_b \tag{62}$$

$$C_{u\,2} = Y_{u22} < \lambda_b, \qquad R_{u\,2} = s(|\tilde{Y}_{12} + \tilde{Y}_{32}|) < 2x\lambda_b < 2\lambda_b \tag{63}$$

$$C_{u\,3} = Y_{u33}, \qquad R_{u\,3} = s(|\tilde{Y}_{13} + \tilde{Y}_{23}|) < 2x\lambda_b < 2\lambda_b. \tag{64}$$

From these estimates we see that circles 1 or 2 can not touch the third (curiously because $7\lambda_b < \lambda_t$) and thus $\lambda_t$ lies certainly inside the third circle, which amounts to the condition

$$|\lambda_t - Y_{u33}| < R_{u\,3}. \tag{65}$$

Combining this inequality with eq. (57) we have, collecting also (56):

$$\lambda_t - 2x\lambda_b \leq |cY_3 + s\tilde{Y}_{33}e^{-i\alpha}| < \lambda_t \tag{66}$$

$$|sY_3 + c\tilde{Y}_{33}e^{-i\alpha}| < \lambda_b. \tag{67}$$

These limits show that to have small $\lambda_b$, either $xY_3$ and $\tilde{Y}_{33}$ are small or there is a cancelation due to $e^{i\alpha} \simeq -1$. The first situation is realized for small $x$, the second has to be realized for larger $x$.

In fact, from the ratio of the two equations $|cY_3 + s\tilde{Y}_{33}\,e^{-i\alpha}| > \lambda_t - 2x\lambda_b$, $|sY_3 + c\tilde{Y}_{33}\,e^{-i\alpha}| < \lambda_b$, we have

$$\left|\frac{x + z\,e^{-i\alpha}}{1 + xz\,e^{-i\alpha}}\right| < \frac{\epsilon}{1 - 2x\epsilon} \simeq \epsilon \tag{68}$$

where $\epsilon = \lambda_b/\lambda_t \simeq 1/30$. and $z = \tilde{Y}_{33}/Y_3$. By allowing $z$ to vary, we find a bound on $\cos\alpha$ as a function of $x$ only:

$$\cos\alpha < \frac{\sqrt{(\epsilon^2 - 1/x)(\epsilon^2 - x)}}{1 - \epsilon^2} \simeq \sqrt{1 - \epsilon^2 x^2 - \epsilon^2/x^2} \tag{69}$$

The following plot shows the range of allowed $\alpha$, for $x > \lambda_b/\lambda_t$:

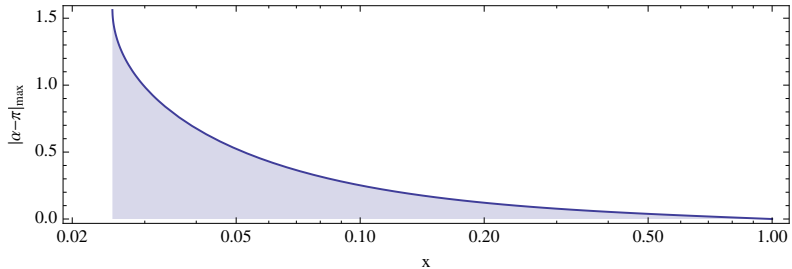

As one can see, as $x$ becomes larger than $\lambda_b/\lambda_t$, the phase $\alpha$ rapidly points in the direction of $\pi$, so that $e^{i\alpha}$ becomes real and the mass matrices are approximately hermitian. The reason why $e^{i\alpha} \simeq -1$ can be understood in the base we have chosen: the same positive elements $\tilde{Y}_{33}$ and $Y_3$ have to give $\lambda_t$ in $Y_u$ but have to interfere destructively in $Y_d$ to produce $\lambda_b$. This requires the phase to be negative to a good approximation.[12]

Summarizing, on one hand for small $x < \lambda_b/\lambda_t$ the phase $\alpha$ is relatively or totally free, but its role for the non-hermiticity of mass matrices is suppressed; on the other, for larger $x$ it is the phase that vanishes. More precisely, with our conventions $\alpha \simeq \pi$. As a result, the up and down mass matrices are always almost hermitian and the mixing matrices are approximately equal.

**Angles and phases.** It is still important to assess the extent to which the right angles can be different from the left ones, because they directly govern the LR contribution to neutral meson mixing and CP violation, and thus to the bound on the LR scale. To this aim, a numerical study of the allowed $V_R$ is most suited, taking into account the present uncertainties on the known quark masses and mixings.

First of all, as a function of $0 < x < 1$ we performed a least-squares fit of the known experimental values (of masses, left mixing angles and CP violating phase). The procedure confirms that the phase $e^{i\alpha}$ is constrained to be near $-1$ for $x > \lambda_b/\lambda_t$ (see figure 6).

We then performed the same fit by demanding also that the Right angles be as different as possible. The deviation of the angles of $V_R$ from the $V_L$ ones is shown in figure 7. As one can see the allowed deviation is quite limited and the left and right angles are similar for any value of $x$.

In the important case of the 'right' Cabibbo angle $\theta_{12\,R}$ we see that it can be deviated at most by 25% by stretching masses and angles by $2\sigma$ (mainly $m_s$ has to be reduced and the Dirac CP phase increased). Thus the bound from $\Delta m_K$ is not reduced appreciably.

Regarding the phases in $V_R$, as discussed above in the case of $\mathcal{P}$ we have only two free complex phases, the spontaneous phase $\alpha$ and a phase of one of the off-diagonal elements of $\tilde{Y}$.

---

[12]We also point out that in the limit $x \to 1$ the mass matrices $M_u$ and $M_d$ tend to be equal, therefore the Yukawa matrices $Y$ and $\tilde{Y}$ hit the perturbativity limit $\sim 3$ around $x \sim 0.8$.

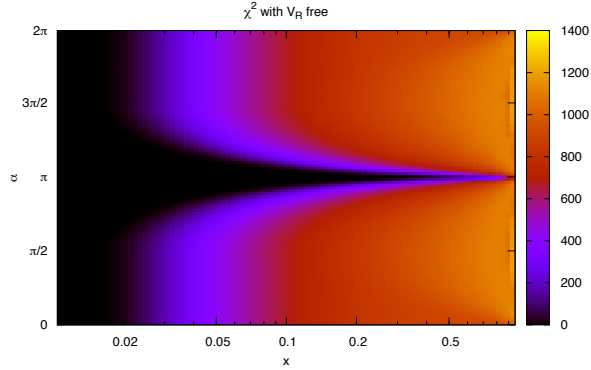

Figure 6: $\chi^2$ in the LR model with $\mathcal{P}$, as a function of the ratio of VEVs $x = v_2/v_1$ and spontaneous phase $\alpha$.

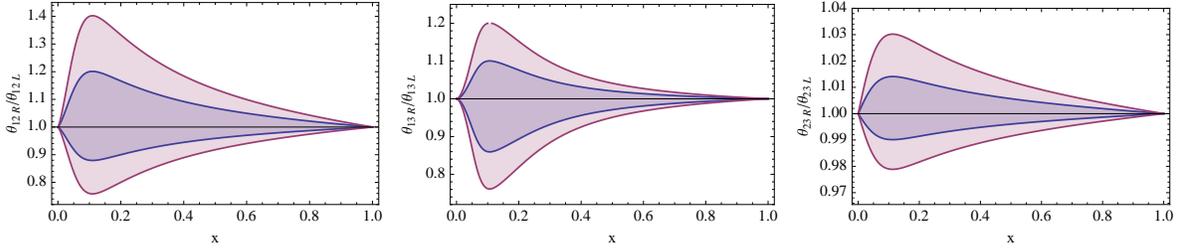

Figure 7: Allowed $V_R$ angles in the LR model with $\mathcal{P}$, as a function of the ratio of VEVs $x$. The shaded zones corresponds to masses and mixing angles deviated at most by $1\sigma$ and $2\sigma$.

All the other phases are determined by these two free parameters. In particular, when $\delta_{CKM}$ is fixed, this leaves only one phase to play with. Therefore all phases are correlated. One finds numerically that $\theta_d - \theta_s$ can be adjusted so as to satisfy the $\epsilon_K$ bound. However, once this is done, the phase $\mathrm{mod}(\theta_b - \theta_s, \pi)$ is also small (order few %) and cannot be moved so as to explain the CP phase in the $B_s$ system (see figure 2).

# References

[1] T. D. Lee and C. N. Yang, Phys. Rev. **104** (1956) 254.

[2] J. C. Pati and A. Salam, Phys. Rev. D **10** (1974) 275. R. N. Mohapatra and J. C. Pati, Phys. Rev. D **11**, 566 (1975); R. N. Mohapatra and J. C. Pati, Phys. Rev. D **11**, 2558 (1975);

[3] G. Senjanović and R. N. Mohapatra, Phys. Rev. D **12**, 1502 (1975); G. Senjanović, Nucl. Phys. B **153** (1979) 334.

[4] G. Senjanović, Theory of neutrino masses and mixings, lectures in the Course CLXX of the International School of Physics Enrico Fermi, Varenna, Italy, June 2008. Published in Measurements of Neutrino Mass, Volume 170 International School of Physics Enrico Fermi, Edited by: F. Ferroni, F. Vissani and C. Brofferio, September 2009.

[5] G. Beall, M. Bander and A. Soni, Phys. Rev. Lett. **48** (1982) 848.

[6] R. N. Mohapatra, G. Senjanović and M. D. Tran, Phys. Rev. D **28**, 546 (1983).

[7] F. J. Gilman and M. H. Reno, Phys. Rev. D **29** (1984) 937; F. J. Gilman and M. H. Reno, Phys. Lett. B **127** (1983) 426.

[8] G. Ecker, W. Grimus and H. Neufeld, Phys. Lett. B **127**, 365 (1983) [Erratum-ibid. B **132**, 467 (1983)].

[9] G. Ecker and W. Grimus, Nucl. Phys. B **258**, 328 (1985).

[10] D. London and D. Wyler, Phys. Lett. B **232** (1989) 503.

[11] K. Kiers, J. Kolb, J. Lee, A. Soni and G. H. Wu, Phys. Rev. D **66** (2002) 095002.

[12] Y. Zhang, H. An, X. Ji and R.N. Mohapatra, Nucl. Phys. B **802** (2008) 247 [arXiv:0712.4218 [hep-ph]].

[13] F. Xu, H. An and X. Ji, JHEP **1003** (2010) 088 [arXiv:0910.2265 [Unknown]].

[14] P. Langacker, U. Sankar, Phys. Rev. D40 (1989) 1569-1585.

[15] G. Ecker, W. Grimus and H. Neufeld, Nucl. Phys. B **229** (1983) 421.

[16] W. Y. Keung and G. Senjanović,

[17] R. N. Mohapatra and G. Senjanović, Phys. Rev. D **23** (1981) 165.

[18] J. Basecq, J. Liu, J. Milutinovic and L. Wolfenstein, Nucl. Phys. B **272** (1986) 145.

[19] M. Bona et al, JHEP **03** (2008) 049; "Status of the Unitarity Triangle Analysis", arXiv:0909.5065.

[20] CKMfitter Group (J. Charles et al.), Eur. Phys. J. **C41**, (2005) 1–131 [hep-ph/0406184], Updated results and plots available at: http://ckmfitter.in2p3.fr;
Vincent Tisserand, on behalf of the CKMfitter group, "CKM fits as of winter 2009 and sensitivity to New Physics", [arXiv:0905.1572v2 [hep-ph]]

[21] S. Herrlich and U. Nierste, Nucl. Phys. B419, 292 (1994); Phys. Rev. D 52, 6505 (1995); Nucl. Phys. B476, 27 (1996);
A. J. Buras, M. Jamin, and P. H. Weisz, Nucl. Phys. B347, 491 (1990).

[22] A.J. Buras, S. Jager, J. Urban, Nucl. Phys. **B605** (2001) 600.

[23] R. Babich, N. Garron, C. Hoelbling, J. Howard, L. Lellouch and C. Rebbi, Phys. Rev. D **74** (2006) 073009.

[24] D. Bećirević, V. Gimenez, G. Martinelli, M. Papinutto and J. Reyes, JHEP **0204** (2002) 025.

[25] H. Wittig. Eur. Phys. J. C33:S890-S894,2004. **Status of lattice calculations of B-meson decays and mixing**

[26] J. P. Silva and L. Wolfenstein, Phys. Rev. D **55** (1997) 5331;
Y. Grossman, Y. Nir and M. P. Worah, Phys. Lett. B **407** (1997) 307;
N. G. Deshpande, B. Dutta and S. Oh, Phys. Rev. Lett. **77** (1996) 4499;
A. Lenz and U. Nierste, JHEP **0706** (2007) 072.

[27] A. J. Buras and D. Guadagnoli, Phys. Rev. D **78** (2008) 033005.

[28] C. Amsler et al. (Particle Data Group), Physics Letters **B667**, 1 (2008)

[29] V. M. Abazov *et al.* [ The D0 Collaboration ], Submitted to: Phys.Rev.D. [arXiv:1005.2757 [hep-ex]].

[30] M. Fabbrichesi, Nucl. Phys. Proc. Suppl. B **99** (2001) 70;
S. Bertolini, AIP Conf. Proc. **618** (2002) 79.

[31] J.M. Frere, J. Galand, A.Le. Yaouanc, L. Oliver, O. P?ne, and J.C. Raynal, Phys. Rev. D **337** (1992) 46.

[32] J. R. Ellis and M. K. Gaillard, Nucl. Phys. B **150** (1979) 141.

[33] E. P. Shabalin, Sov. J. Nucl. Phys. **32** (1980) 228 [Yad. Fiz. **32** (1980) 443].

[34] P. Minkowski, Phys. Lett. B **67** (1977) 421; T. Yanagida, proceedings of the *Workshop on Unified Theories and Baryon Number in the Universe*, Tsukuba, 1979, eds. A. Sawada, A. Sugamoto, KEK Report No. 79-18, Tsukuba; S. Glashow, in *Quarks and Leptons, Cargèse 1979*, eds. M. Lévy. et al., (Plenum, 1980, New York); M. Gell-Mann, P. Ramond, R. Slansky, proceedings of the *Supergravity Stony Brook Workshop*, New York, 1979, eds. P. Van Niewenhuizen, D. Freeman (North-Holland, Amsterdam); R. Mohapatra, G. Senjanović, Phys.Rev.Lett. **44** (1980) 912

[35] A. Ferrari *et al.*, Phys. Rev. D **62** (2000) 013001.

[36] S. N. Gninenko, M. M. Kirsanov, N. V. Krasnikov and V. A. Matveev, Phys. Atom. Nucl. **70** (2007) 441.

[37] V. Bansal, arXiv:0910.2215 [Unknown].

[38] A. Datta and A. Pilaftsis, Phys. Lett. B **278**, 162 (1992).

[39] J. Kersten and A. Y. Smirnov, Phys. Rev. D **76** (2007) 073005.

[40] T. Han, B. Mukhopadhyaya, Z. Si and K. Wang, Phys. Rev. D **76**, 075013 (2007) [arXiv:0706.0441 [hep-ph]].

[41] M. Kadastik, M. Raidal and L. Rebane, Phys. Rev. D **77**, 115023 (2008).

[42] G. Azuelos, K. Benslama and J. Ferland, J. Phys. G **32** (2006) 73.

[43] A. G. Akeroyd and M. Aoki, Phys. Rev. D **72** (2005) 035011.

[44] J. Garayoa and T. Schwetz, JHEP **0803** (2008) 009.

[45] C. S. Aulakh, A. Melfo and G. Senjanović, Phys. Rev. D **57**, 4174 (1998).

[46] Z. Chacko and R. N. Mohapatra, Phys. Rev. D **58**, 015003 (1998).

[47] C. S. Aulakh, A. Melfo, A. Rašin and G. Senjanović, Phys. Rev. D **58**, 115007 (1998).

[48] M. Nemevšek, F. Nesti, G. Senjanović, V. Tello, *to appear*.

[49] C. Ankenbrandt *et al.*, arXiv:physics/0611124.

[50] http://j-parc.jp/NuclPart/pac_0701/pdf/P21-LOI.pdf;
http://j-parc.jp/NuclPart/pac_0606/pdf/p20-Kuno.pdf

[51] A. G. Akeroyd, M. Aoki and Y. Okada, Phys. Rev. D **76** (2007) 013004.

[52] B. Bajc, M. Nemevšek and G. Senjanović, Phys. Lett. B **684** (2010) 231.

[53] C.S. Lim and T. Inami, Prog. Theor. Phys. 67, 1569 (1982); M. L. Swartz, Phys. Rev. D 40, 1521 (1989);
V. Cirigliano, A. Kurylov, M.J. Ramsey-Musolf, and P. Vogel, Phys. Rev. D 70, 075007 (2004).

[54] S. Gershgorin, "Über die Abgrenzung der Eigenwaerte einer Matrix", Izv. Akad. Nauk. USSR Otd. Fiz.-Mat. Nauk. 7, 749–754, 1931.

[55] T. Inami and C. S. Lim, Prog. Theor. Phys. **65** (1981) 297 [Erratum-ibid. **65** (1981) 1772].